\documentclass{aa}

\usepackage{pifont}

\usepackage{tikz}
\newcommand{\tixzxmark}{%
\tikz[scale=0.23] {
    \draw[line width=0.7,line cap=round] (0,0) to [bend left=6] (1,1);
    \draw[line width=0.7,line cap=round] (0.2,0.95) to [bend right=3] (0.8,0.05);
}}
\usepackage{bm}
\usepackage{soul}
\usepackage{multirow}
\usepackage{longtable,tabu}

\newcommand{\be}{\begin{equation}}
\newcommand{\ee}{\end{equation}}
\newcommand{\bea}{\begin{eqnarray}}
\newcommand{\eea}{\end{eqnarray}}

\usepackage{amsmath}

\usepackage{url}

\usepackage{xcolor}

\definecolor{color1}{HTML}{440154}
\definecolor{color2}{HTML}{481568}
\definecolor{color3}{HTML}{482677}

\definecolor{color18}{HTML}{B8DE29}
\definecolor{color19}{HTML}{DCE318}
\definecolor{color20}{HTML}{FDE725}
\graphicspath{{./}{figures/}}

\begin{document}

\title{Unified neutron star equations of state calibrated to nuclear properties}

\author{Tuhin Malik\thanks{tuhin.malik@uc.pt} \and Helena Pais\thanks{hpais@uc.pt} \and Constan\c ca Provid\^encia\thanks{cp@uc.pt}}

\institute{CFisUC, University of Coimbra, P-3004-516 Coimbra, Portugal.} 

\date{Received xxx Accepted xxx}


 \abstract
  {Recently, in \cite{Malik:2023mnx}, a dataset of several equations of state (EoS) for purely nucleonic stellar matter based on a non-linear relativistic-mean field (RMF) model prescription, and constrained to properties of nuclear matter, to state-of-the-art chiral effective field theory calculations for low-density neutron matter, and to astrophysical data, were proposed.} 
{In this work, twenty one unified neutron star EoS were chosen from that dataset, in such a way that a large range of values of the slope of the symmetry energy at saturation is covered.  Several quantities are calculated and discussed, such as the the proton fraction and the direct Urca behavior, the density dependence of the  speed of sound and the trace anomaly, the crust-core transition properties,  the compatibility with  astrophysical observations, and the neutron matter properties from chiral effective field theory ($\chi$EFT) calculations and pQCD constraints.}   
{We construct unified EoS, where the outer crust is given by the BSk22 functional, and the inner crust is calculated from a compressible liquid drop (CLD) approximation. The core is purely nucleonic, made of protons, neutrons, electrons and muons, under charge neutrality and in $\beta-$equilibrium conditions.}
{The correlation of the slope of the symmetry energy at saturation with the crust-core transition density and proton fraction is analysed, and equations that translate these relations are proposed.  Moreover, the spectral representation for all the EOS according to the format proposed in \cite{Lindblom:2010bb} is given, which is a convenient representation to study quasi-periodic oscillations with realistic EOS. It is shown that several of these EoS have in the center of the most massive NS a speed of sound squared of the order of $\lesssim 0.5$.  Most of the EoS predict a maximum central density of the order of about 6 times the nuclear saturation density. Three of the EoS satisfy all of the constraints imposed.  All these EoS will be made available in the CompOSE platform. }
   {}


\authorrunning{Malik et al.}

\maketitle

\section{Introduction} \label{sec:intro}

The macro and microscopic properties of neutron stars (NS) are still under study nowadays. However, with the present class of instrumentation, astrophysical measurements of both mass and radius of pulsars \citep{Riley:2019yda,Miller:2019cac,Riley:2021pdl,Miller:2021qha,Raaijmakers:2021uju} and also the detection of gravitational wave signals from the merger of two NS, such as GW170817 \citep{TheLIGOScientific:2017qsa} or the GW190425 \citep{LIGOScientific:2020aai}, have been possible, and this has allowed the nuclear physics community, working on the equation of state (EoS) for stellar matter, to further constraint their calculations. Microscopic calculations, such as \textit{ab-initio} neutron matter calculations from $\chi$EFT models \citep{Hebeler2013,Drischler:2017wtt,Drischler:2020yad} have also been used to constraint the nuclear matter EoS in the low-density regime, i.e. at sub-saturation densities.

In the inner crust, heavy clusters (\textit{nuclear pasta}) are expected to form due to the competition between the strong and Coulomb forces \citep{Ravenhall:1983uh}. These geometries will have an effect on the transport properties and cooling of the star, because of the possible modification of the neutrino mean-free path \citep{Arcones:2008kv}. This layer of the star plays a big role in the determination of the radius of the star (see e.g \cite{Pais:2016xiu} or \cite{Fortin:2016hny}), since this quantity is affected by the determination of the crust-core transition.
However, these nuclear clusters that form may pose a problem for its numerical calculation, so polytropic mechanisms to mimic the crust of the star,  such as the one used in \cite{Carriere:2002bx,Malik:2022jqc}, are often used.

Another quantity associated with the inner crust, and that has an effect, not only on $R$, but also on the slope of the symmetry energy, $L$, is the density at which the nuclear clusters melt, and the core, constituted by a homogeneous gas of protons, neutrons, and electrons and muons, starts. This is called the crust-core transition density, $\rho_t$. Previous studies (see e.g. \cite{Pais:2016xiu,Ducoin:2010as,Newton:2011dw,Providencia:2013dsa}) have observed an anti-correlation between this quantity and $L$. Other studies have also tested other correlations (and combination of correlations) between nuclear matter parameters and EoS properties, such as $\rho_t$ or the associated pressure, $P_t$ (see e.g. \cite{Ducoin:2011fy} or \cite{Alam:2016cli}). 

Besides the numerical computation of the inner crust, the lack of knowledge of the nuclear EoS at high densities together with the fact that the whole range of the star spans several orders of magnitude in density, makes the computation of a fully unified EoS a difficult task. Different prescriptions have been then proposed to contour this problem, such as metamodeling techniques \citep{Margueron:2017eqc,Margueron:2017lup,Ferreira:2019bgy,Xie:2019sqb,Xie:2020tdo,Ferreira:2021pni,Thi:2021jhz}, that make a series expansion of the energy per nucleon to generate a realistic EoS,  or non-parametric methods \citep{Landry:2018prl,Essick:2019ldf,Gorda:2022jvk,Zhou:2023cfs}, that infer the EoS directly from GW data. However, the composition is not determined within these kind of methods.

The density-dependence of the symmetry energy and its effect on the macroscopic properties of the star has been explored in several works (see e.g. \cite{Pais:2016xiu,Cavagnoli:2011ft,Providencia:2012rx,Bao:2014lqa,Ji:2019hxe}), where relativistic mean-field models with non-linear meson terms have been used. In particular, the larger the $\Lambda_{\omega}$ parameter associated with the vector-isovector $\omega\rho$ mixing term is, the smaller the symmetry energy at saturation \citep{Cavagnoli:2011ft}. Moreover, it has been shown that there exists a correlation between the non-linear vector meson $\omega$ term  ($\omega^4$ term) and the macroscopic properties of the star, such as the radius \citep{Fattoyev:2010rx}. This term seems to also affect the speed of sound, and a correlation between this term and the symmetry energy, has also been observed \citep{Malik:2023mnx}. The symmetry energy is known to play an important role in the composition of the star, and, consequently, on the nucleonic direct Urca processes \citep{Yakovlev:2000jp,Yakovlev:2004iq,Fortin:2020qin,10.3389/fspas.2019.00013}: if the symmetry energy is very low, very asymmetric matter is favoured and this will not allow for such kind of processes in the interior of NS (see e.g. \cite{Providencia:2023rxc} and references therein).

The behaviour of the speed of sound has been also recently used to check when other, exotic, degrees of freedom set in, such as hyperons \citep{Malik:2022jqc}, or even deconfined quark matter \citep{Annala2019,Altiparmak:2022bke,Somasundaram:2022ztm}. Another quantity that has been used to test deconfinement is the trace anomaly, renormalized in \cite{Fujimoto:2022xhv} as $\Delta=1/3-P/\epsilon$, meaning that it should tend to zero in the conformal limit. In \cite{Annala:2023cwx}, another related quantity was proposed, based also on the speed of sound, $d_c=\sqrt{\Delta^2+ {\Delta}^{'2}}$, where  $\Delta'= c_s^2 \, \left(1/\gamma-1\right)$  is the logarithmic derivative of $\Delta$ with respect to the energy density, which should also approach zero in the conformal limit. 

In this work, we present 21 unified EOS based on the relativistic mean field approach used in \cite{Malik:2023mnx} including the constraints imposed in this work, and discuss their properties. The criterium to select the  EOS is the slope of the symmetry energy at saturation, in particular, we consider values between 22 MeV and 70 MeV. The inner crust is built considering a compressible liquid drop approximation \citep{Pais15}. Regarding the crust-core transition properties, we compare the values obtained from such a calculation with the ones obtained from a dynamical spinodal one \citep{Pais:2016xiu}.  
Neutron star properties, such as the mass-radius curve, the tidal deformability, the crust-core transition density, the proton fraction, the speed of sound and the trace anomaly, among others, are given.  A comparison between the performance of the different EOS and the NS observations and the neutron matter EOS  from a $\chi$EFT calculation is discussed.  We also give the spectral representation for all the EOS according to the format proposed in \cite{Lindblom:2010bb}, which is a convenient representation to study quasi-periodic oscillations with realistic EOS. A discussion of the error introduced when the unified inner crust is not used, as in \cite{Malik:2023mnx}, is  also discussed.

\section{Formalism \label{formalism}}

The twenty-one EoS are obtained within an {RMF} approach. The Lagrangian density that generates these models is written in terms of the fermionic fields that describe the nucleons, and the mesonic fields, the scalar isoscalar $\sigma$ field, the vector isoscalar $\omega$ field, and the vector isovector $\rho$ field, responsible for the description of the nuclear interaction. In the present formalism, non-linear meson terms are included to allow for a realistic modulation of the density dependence of both the symmetric nuclear matter and the symmetry energy. The role of these terms has been discussed in several seminal works \cite{Boguta1977,Horowitz:2000xj,Mueller:1996pm,Sugahara:1993wz}, and explored within a Bayesian inference study  in \cite{Malik:2023mnx}. The parameters are listed in Table \ref{tab:1}.

\begin{table*}
\caption{The parameters for all 21 models obtained from the Lagrangian density \ref{lag}. Specifically, B and C are $b \times 10^3$ and $c \times 10^3$, respectively. The nucleon, $\omega$ meson, $\sigma$ meson, and $\varrho$ meson masses considered are 939, 782.5, 500, and 763 MeV, respectively. \label{tab:1} } 
\setlength{\tabcolsep}{12.pt}
      \renewcommand{\arraystretch}{1.1}
\begin{center}
\begin{tabular}{cccccccc}
\hline
\hline
EOS &         $g_\sigma$ &         $g_\omega$ &         $g_\rho$ &         $B$ &         $C$ &        $\xi$ &       $\Lambda_\omega$ \\
\hline
EOS1   &  10.411847 &  13.219028 &  11.180337 &  2.541001 & -3.586261 &  0.000845 &  0.027999 \\
EOS2   &  10.485889 &  13.447123 &   9.640716 &  2.204058 & -2.045914 &  0.015365 &  0.030662 \\
EOS3   &  10.547924 &  13.478944 &  10.741298 &  2.345282 & -2.397921 &  0.013874 &  0.034780 \\
EOS4   &  10.764953 &  13.892817 &  16.316137 &  1.931926 & -1.349144 &  0.017592 &  0.029602 \\
EOS5   &  11.150279 &  14.420375 &  13.806001 &  2.036239 & -1.635468 &  0.018019 &  0.037600 \\
EOS6  &   8.027986 &   9.144665 &  13.334730 &  6.478531 & -4.162544 &  0.001258 &  0.100686 \\
EOS7  &   8.219347 &   9.263474 &  13.094357 &  7.304950 & -4.472765 &  0.000789 &  0.089162 \\
EOS8 &   8.637377 &  10.348224 &  11.228904 &  3.910898 & -2.158740 &  0.001478 &  0.078386 \\
EOS9 &   8.666023 &  10.123372 &  14.838137 &  5.068521 & -3.068444 &  0.000005 &  0.072946 \\
EOS10 &   8.674249 &  10.196514 &  14.188737 &  4.605767 & -2.108763 &  0.000754 &  0.069895 \\
EOS11 &   8.695491 &  10.431351 &   9.821776 &  3.975509 & -2.615425 &  0.006394 &  0.039323 \\
EOS12 &   8.720440 &  10.481072 &  12.378912 &  3.597999 & -1.429640 &  0.002009 &  0.050647 \\
EOS13 &   8.789344 &  10.482588 &   9.304529 &  4.597276 & -4.641323 &  0.001198 &  0.031954 \\
EOS14 &   8.885870 &  11.185655 &  11.060177 &  1.921115 & -0.614536 &  0.006671 &  0.051142 \\
EOS15 &   8.973194 &  10.876543 &  14.515228 &  3.783637 & -3.414168 &  0.004491 &  0.055611 \\
EOS16 &   9.123130 &  11.557699 &  11.708959 &  1.875115 & -0.441126 &  0.012244 &  0.036703 \\
EOS17 &   9.219420 &  11.126923 &   9.877649 &  4.039626 & -4.268039 &  0.003467 &  0.035905 \\
EOS18 &   9.220247 &  11.170082 &  11.087122 &  4.036904 & -4.553914 &  0.003810 &  0.040742 \\
EOS19 &   9.295923 &  11.527999 &  11.103228 &  3.135381 & -3.390132 &  0.007847 &  0.039448 \\
EOS20 &   9.554944 &  11.640795 &  14.692091 &  3.887915 & -4.661381 &  0.003635 &  0.043001 \\
EOS21 &   9.608190 &  11.957725 &  12.191950 &  3.117923 & -4.098400 &  0.000255 &  0.058744 \\
\hline
\hline
\end{tabular} 
\end{center}
\end{table*}

\subsection{The model}

 The Lagrangian describing the nuclear matter system is given by
\begin{equation}
\label{lag}
  \mathcal{L}=   \mathcal{L}_N+ \mathcal{L}_M+ \mathcal{L}_{NL}
\end{equation} 
with
\begin{equation}
\begin{aligned}
\mathcal{L}_{N}=& \bar{\Psi}\Big[\gamma^{\mu}\left(i \partial_{\mu}-g_{\omega} \omega_{\mu}-
g_{\varrho} {\boldsymbol{t}} \cdot \boldsymbol{\varrho}_{\mu}\right) \\
&-\left(m-g_{\sigma} \sigma\right)\Big] \Psi \\
\mathcal{L}_{M}=& \frac{1}{2}\left[\partial_{\mu} \sigma \partial^{\mu} \sigma-m_{\sigma}^{2} \sigma^{2} \right] \\
&-\frac{1}{4} F_{\mu \nu}^{(\omega)} F^{(\omega) \mu \nu} 
+\frac{1}{2}m_{\omega}^{2} \omega_{\mu} \omega^{\mu} \nonumber\\
&-\frac{1}{4} \boldsymbol{F}_{\mu \nu}^{(\varrho)} \cdot \boldsymbol{F}^{(\varrho) \mu \nu} 
+ \frac{1}{2} m_{\varrho}^{2} \boldsymbol{\varrho}_{\mu} \cdot \boldsymbol{\varrho}^{\mu}.\\
    			\mathcal{L}_{NL}=&-\frac{1}{3} b ~m~ g_\sigma^3 (\sigma)^{3}-\frac{1}{4} c g_\sigma^4 (\sigma)^{4}+\frac{\xi}{4!}g_{\omega}^4(\omega_{\mu}\omega^{\mu})^{2} \nonumber\\&+\Lambda_{\omega}g_{\varrho}^{2}\boldsymbol{\varrho}_{\mu} \cdot \boldsymbol{\varrho}^{\mu} g_{\omega}^{2}\omega_{\mu}\omega^{\mu},
\end{aligned}
\label{lagrangian}
\end{equation}
where $\Psi$ are the   Dirac spinors that describe the nucleons, protons and neutrons,  with a  bare mass $m$. In the above expression, 
$\gamma^\mu $  are the Dirac matrices,  $\boldsymbol{t}$ is the isospin operator. We have introduced  the following vector meson  tensors   $F^{(\omega, \varrho)\mu \nu} = \partial^ \mu A^{(\omega, \varrho)\nu} -\partial^ \nu A^{(\omega, \varrho) \mu}$.  The parameters $g_{\sigma}$, $g_{\omega}$ and $g_{\varrho}$ designate the couplings  of the nucleons to the meson fields $\sigma$, $\omega$ and $\varrho$ with masses, respectively,  $m_\sigma$, $m_\omega$ and $m_\varrho$. 

The coupling parameters of the non-linear terms
$b,\, c,$ $\xi$ and  $\Lambda_{\omega}$ together with the couplings $g_i$  have  been calculated imposing a set of constraints which will be next discussed. The role  of the non-linear terms is briefly summarized:  the non-linear $\sigma$ terms introduced by \cite{Boguta1977} control the nuclear matter incompressibility at saturation; the $\omega^4$ term  was introduced in \cite{Sugahara:1993wz} and softens the EoS at high densities; the $\omega\rho$ term allows to control the density dependence of the symmetry energy.
The meson-fields are taken as static,  classical fields and replaced by their expectation values determined from the equations
		\begin{eqnarray}
			{\sigma}&=& \frac{g_{\sigma}}{m_{\sigma,{\rm eff}}^{2}}\sum_{i} \rho^s_i\label{sigma}\\
			{\omega} &=&\frac{g_{\omega}}{m_{\omega,{\rm eff}}^{2}} \sum_{i} \rho_i \label{omega}\\
			{\varrho} &=&\frac{g_{\varrho}}{m_{\varrho,{\rm eff}}^{2}}\sum_{i} I_{3} \rho_i, \label{rho}
		\end{eqnarray}
 with $\rho^s_i$ and $\rho_i$, respectively, the scalar density and the number density of nucleon $i$, and
 \begin{eqnarray}
   m_{\sigma,{\rm eff}}^{2}&=& m_{\sigma}^{2}+{ b ~m~g_\sigma^3}{\sigma}+{c g_\sigma^4}{\sigma}^{2} \label{ms} \\ 
    m_{\omega,{\rm eff}}^{2}&=& m_{\omega}^{2}+ \frac{\xi}{3!}g_{\omega}^{4}{\omega}^{2} +2\Lambda_{\omega}g_{\varrho}^{2}g_{\omega}^{2}{\varrho}^{2}\label{mw}\\
    m_{\varrho,{\rm eff}}^{2}&=&m_{\varrho}^{2}+2\Lambda_{\omega}g_{\omega}^{2}g_{\varrho}^{2}{\omega}^{2}. \label{mr}
 \end{eqnarray}
 
In order to describe $\beta$-equilibrium matter, electrons and muons are introduced in the Lagrangian density
$${\cal L}_l=\bar{\psi}_l \gamma^{\mu}\left(i \partial_{\mu}-m_l\right) \psi_l$$
Electric charge neutrality and $\beta$-equilibrium are imposed, defined by, respectively, the relations
$$\rho_p=\rho_e+\rho_\mu,$$
$$\mu_n=\mu_p+\mu_e, \quad \mu_\mu=\mu_e.
$$
In order, to determine the NS properties, in particular, to integrate the Tolmann-Oppenheimer-Volkoff equations \cite{Tolman:1939jz,Oppenheimer:1939ne} to obtain the mass and radius of spherical static NS,  the pressure and energy density must be calculated. The energy density is given by
			\begin{equation}
			\begin{aligned}
			\epsilon &= \sum_{i=n,p,e,\mu}\frac{1}{2\pi^2}\int_0^{k_{Fi}} \sqrt{k^2+m_i^*}\, k^2\, dk \\
			&+ \frac{1}{2}m_{\sigma}^{2}{\sigma}^{2}+\frac{1}{2}m_{\omega}^{2}{\omega}^{2}+\frac{1}{2}m_{\varrho}^{2}{\varrho}^{2}\\
			&+ \frac{b\,m}{3}(g_{\sigma}{\sigma})^{3}+\frac{c}{4}(g_{\sigma}{\sigma})^{4}+\frac{\xi}{8}(g_{\omega}{\omega})^{4} + 3\Lambda_{\omega}(g_{\varrho}g_{\omega}{\varrho}{\omega})^{2},
			\end{aligned}
			\end{equation}
   where $m_i^*=m_i-g_s \sigma$ for protons and neutrons and $m_i^*=m_i$ for electrons and muons, and $k_{Fi}$ is the Fermi moment of particle $i$. 
The pressure is determined from the thermodynamic relation
			\begin{equation}
			P = \sum_{i}\mu_{i}\rho_{i}-\epsilon.
			\end{equation}

\subsection{The outer crust}

The outer crust region of the star is composed by a neutral lattice of nuclei and electrons. Even though there are several EoS in the literature for this layer of the star, in what concerns the macroscopic properties of the star, such as the mass and radius, the outer crust does not affect any of those, and, in principle, any EoS can be used. In the CompOSE website, a public free online database collection of EoS for compact objects, we can find several of them.
For this region, we chose the BSk22 EoS \citep{Pearson18}, taken from the CompOSE website \citep{Typel2015-compose,Typel2022-compose,compose}. This EoS was calculated using the Hartree-Fock-Bogoliubov atomic mass table HFB-22, and the 2016 Atomic Mass Evaluation \citep{Audi17}, when the former masses were not available (see ~\cite{composeBSk22} and references therein).

 \subsection{The inner crust}
 \label{cld}

 In the inner crust, heavy clusters are expected to form. Due to the competition between the strong and Coulomb forces, they form geometrical structures, being called the nuclear pasta phase due to their resemblance to the Italian food. These clusters may affect the cooling of the proton-neutron star, as the neutrino mean-free path may be affected, and they may also have consequence on the dynamics of the object, namely on the transport properties. Eventually these geometries will melt, and the core of the star is reached. This is denoted the crust-core transition.

 In this work, this region of the star is calculated within the compressible liquid drop (CLD) model \citep{Pais15}, considering $\beta-$equilibrium and zero temperature stellar matter. In this approximation, the system is composed of two distinct regions, the low-density one, denoted by $II$, i.e. the gas, composed of free neutrons, protons, and electrons, and the high-density one, denoted by $I$, the clusters.  The total energy density of the system, the conditions for electrical neutrality, and $\beta-$equilibrium, are given respectively by

 \begin{eqnarray}
F&=&fF^I + (1-f)F^{II} + F_e +\varepsilon_{surf} + \varepsilon_{Coul} , \label{totalfree}\\
\rho_p&=&\rho_e=f\rho_p^I+(1-f)\rho_p^{II} \, , \\
\mu_n&=&\mu_p+\mu_e \, , 
\end{eqnarray}
where $F^i, i=I,II,$ is the free energy density of the homogeneous neutral nuclear matter, and the Coulomb and surface terms are given respectively by

\begin{eqnarray}
\varepsilon_{Coul}&=&2\alpha e^2 \pi \Phi R_d^2 \left(\rho_p^I-\rho_p^{II}\right)^2 \, , \\
\varepsilon_{surf}&=&\alpha D \sigma / R_d  \, .
\end{eqnarray}
 
To find the stable solutions, the total energy is minimized against the variables of the system: the proton and neutron densities, in both phases, $I$ and $II$, the radius of the geometry, $r_d$, and the fraction of phase $I$, $f$. From this minimization, we obtain the following equilibrium relations:

\begin{eqnarray}
\varepsilon_{surf}&=& 2\varepsilon_{coul}  \, , \nonumber \\
\mu_n^I&=&\mu_n^{II}   \, ,  \nonumber \\
\mu_p^I&=&\mu_p^{II}-\frac{2\varepsilon_{coul}}{f(1-f)(\rho_p^I-\rho_p^{II})}  \nonumber  \, , \\
P^{I}&=&P^{II}-\frac{2\varepsilon_{Coul}}{(\rho_p^I-\rho_p^{II})}\left(\frac{\rho_p^I}{f}+\frac{\rho_p^{II}}{(1-f)}\right) \nonumber \\
&+&\varepsilon_{Coul}\left(\frac{3}{\alpha}\frac{\partial \alpha}{\partial f}+\frac{1}{\Phi}\frac{\partial\Phi}{\partial f}\right) \, .
\end{eqnarray}
with the radius of the geometry given by
\begin{eqnarray}
R_d&=&\left(\dfrac{D \sigma}{4\pi e^2\Phi\left(\rho_p^I-\rho_p^{II}\right)^2}\right)^{1/3}    
\end{eqnarray}
In these equations, $\alpha=f$ for droplets, rods, slabs and $\alpha=1-f$ for tubes and bubbles, $\sigma$ is the surface energy coefficient, $D$ is the dimension of the geometry (3 for droplets), and $\Phi$ is given by 
\begin{eqnarray}
\Phi=\left\lbrace\begin{array}{c}
\left(\frac{2-D \alpha^{1-2/D}}{D-2}+\alpha\right)\frac{1}{D+2}, D=1,3 \\
\frac{\alpha-1-\ln \alpha}{D+2}, D=2 \quad .
\end{array} \right. 
\end{eqnarray}

\section{Results}

A collection of 21 equations of state  based on Relativistic Mean Field (RMF) models, featuring a range of nuclear saturation properties, was created to explore the characteristics of neutron stars. The significance and implications of these 21 EOS models as well as their intricate properties will be discussed in the following.

A significant feature of neutron stars is their inner crust, which consists of non-homogeneous clustered matter, where we used a CLD calculation for the inner crust. Since this calculation is not fully self-consistent as the surface tension was chosen to be the same for all the models, we also perform a  dynamical spinodal calculation to calculate the transition density. These two different approaches will be discussed in the following.   
In all the figures, we have highlighted the three EoS, EoS 8, 18, and 19, that have passed all the constraints discussed in Section~\ref{constraints}.

\subsection{Properties of the EoS}

\begin{table*}
\caption{The following nuclear saturation properties are listed for all EOS models: saturation density (fm$^{-3}$),  binding energy per nucleon $\epsilon_0$, incompressibility of nuclear matter $K_0$, skewness $Q_0$, kurtosis $Z_0$, symmetry energy $J_{\rm sym,0}$, slope $L_{\rm sym,0}$, curvature $K_{\rm sym,0}$, skewness $Q_{\rm sym,0}$ and kurtosis $Z_{\rm sym,0}$ of symmetry energy, all in MeV,  are evaluated at nuclear saturation density $\rho_0$. Furthermore, we list the transition density $\rho_t$ (fm$^{-3}$), $\beta$- equilibrium pressure $P_t$ (MeV.fm$^{-3}$), and proton fraction $y_p$ at the transition density calculated with both the CLD and dynamic spinodal approach.} 
\label{tab:2}
\setlength{\tabcolsep}{3.0pt}
      \renewcommand{\arraystretch}{1.1}
\begin{center}      
\begin{tabular}{ccccccccccccccccc}
\hline \hline
\multirow{2}{*}{model} & \multirow{2}{*}{$\rho_0$} & \multirow{2}{*}{$\epsilon_0$} & \multirow{2}{*}{$K_0$} & \multirow{2}{*}{$Q_0$} & \multirow{2}{*}{$Z_0$} & \multirow{2}{*}{$J_{\rm sym,0}$} & \multirow{2}{*}{$L_{\rm sym,0}$} & \multirow{2}{*}{$K_{\rm sym,0}$} & \multirow{2}{*}{$Q_{\rm sym,0}$} & \multirow{2}{*}{$Z_{\rm sym,0}$} & \multicolumn{3}{c}{CLD}    & \multicolumn{3}{c}{Dynamical spinodal} \\ \cline{12-17} 
                       &                           &                               &                        &                        &                        &                                  &                                  &                                  &                                  &                                  & $\rho_t$ & $P_t$  & $y_p$  & $\rho_t$     & $P_t$      & $y_p$      \\ \hline
EOS1                   & 0.155                     & -16.08                        & 177                    & -74                    & 18944                  & 33                               & 64                               & 77                               & 1741                             & -17088                           & 0.0781   & 0.5446 & 0.0349 & 0.0773       & 0.5204     & 0.0341     \\
EOS2                   & 0.161                     & -16.11                        & 206                    & 279                    & 12222                  & 31                               & 70                               & 21                               & 551                              & -13709                           & 0.0672   & 0.3599 & 0.0225 &  0.0638	& 0.3028	& 0.0202
   \\
EOS3                   & 0.163                     & -16.15                        & 192                    & 375                    & 15288                  & 32                               & 66                               & 59                               & 552                              & -17990                           & 0.0726   & 0.4173 & 0.0287 & 0.0765       & 0.4475     & 0.0290     \\
EOS4                   & 0.149                     & -16.11                        & 209                    & 87                     & 8883                   & 35                               & 47                               & 79                               & 402                              & -16785                           & 0.0944   & 0.4781 & 0.0673 & 0.0849       & 0.3202     & 0.0648     \\
EOS5                   & 0.154                     & -15.72                        & 190                    & 614                    & 12505                  & 32                               & 60                               & 98                               & -68                              & -14841                           & 0.0726   & 0.3180 & 0.0407 & 0.0741       & 0.2959     & 0.0391     \\
EOS6                   & 0.169                     & -16.4                         & 252                    & -451                   & 2181                   & 33                               & 26                               & -68                              & 1492                             & -18264                           & 0.127    & 0.7313 & 0.0613 & 0.1182       & 0.5178     & 0.0615     \\
EOS7                   & 0.15                      & -15.67                        & 231                    & -472                   & 2195                   & 32                               & 28                               & -128                             & 1653                             & -14718                           & 0.1107   & 0.7956 & 0.0586 & 0.1041       & 0.6686     & 0.0572     \\
{\bf EOS8}                   & 0.151                     & -15.99                        & 267                    & -332                   & 1654                   & 29                               & 36                               & -77                              & 1210                             & -11321                           & 0.0944   & 0.4704 & 0.0373 & 0.0958       & 0.4456     & 0.0363     \\
EOS9                   & 0.144                     & -16.28                        & 254                    & -420                   & 2066                   & 33                               & 22                               & -72                              & 1601                             & -19302                           & 0.1162   & 0.6369 & 0.0676 & 0.1048       & 0.4110     & 0.0687     \\
EOS10                  & 0.144                     & -16.33                        & 262                    & -401                   & 1766                   & 32                               & 25                               & -89                              & 1606                             & -17661                           & 0.1107   & 0.6291 & 0.0634 & 0.1018       & 0.4570     & 0.0626     \\
EOS11                  & 0.157                     & -16.24                        & 260                    & -400                   & 2173                   & 32                               & 57                               & -134                             & 864                              & -2126                            & 0.0835   & 0.5762 & 0.0303 & 0.0862       & 0.5999     & 0.0305     \\
EOS12                  & 0.147                     & -16.22                        & 276                    & -330                   & 1388                   & 33                               & 39                               & -137                             & 1529                             & -11075                           & 0.0998   & 0.7261 & 0.0528 & 0.0966       & 0.6449     & 0.0508     \\
EOS13                  & 0.156                     & -16.15                        & 238                    & -423                   & 3038                   & 32                               & 63                               & -124                             & 632                              & 173                              & 0.0781   & 0.5144 & 0.0252 & 0.0799       & 0.5365     & 0.0256     \\
EOS14                  & 0.162                     & -16.18                        & 316                    & -100                   & 808                    & 31                               & 46                               & -59                              & 1154                             & -12403                           & 0.1053   & 0.5870 & 0.0407 & 0.1013       & 0.5183     & 0.0392     \\
EOS15                  & 0.153                     & -16.16                        & 245                    & -388                   & 2900                   & 34                               & 29                               & -33                              & 1478                             & -19325                           & 0.1162   & 0.6430 & 0.0654 & 0.1055       & 0.4299     & 0.0655     \\
EOS16                  & 0.157                     & -15.78                        & 290                    & -211                   & 1283                   & 34                               & 50                               & -87                              & 1272                             & -12007                           & 0.0998   & 0.7242 & 0.0479 & 0.0970       & 0.6518     & 0.0460     \\
EOS17                  & 0.152                     & -16.62                        & 233                    & -470                   & 3829                   & 31                               & 56                               & -107                             & 978                              & -3122                            & 0.0781   & 0.5039 & 0.0282 & 0.0799       & 0.5186     & 0.0283     \\
{\bf EOS18}                  & 0.155                     & -16.31                        & 221                    & -491                   & 4260                   & 32                               & 49                               & -92                              & 1388                             & -8841                            & 0.0889   & 0.6671 & 0.0394 & 0.0875       & 0.6263     & 0.0381     \\
{\bf EOS19}                  & 0.165                     & -16.24                        & 230                    & -332                   & 5616                   & 33                               & 53                               & -48                              & 1332                             & -12465                           & 0.0944   & 0.6914 & 0.0409 & 0.0934       & 0.6506     & 0.0396     \\
EOS20                  & 0.147                     & -16.18                        & 199                    & -567                   & 4651                   & 34                               & 32                               & -19                              & 1628                             & -18997                           & 0.1053   & 0.6606 & 0.0657 & 0.0976       & 0.5082     & 0.0649     \\
EOS21                  & 0.156                     & -16.12                        & 216                    & -339                   & 6785                   & 29                               & 42                               & 55                               & 1146                             & -14120                           & 0.0944   & 0.3955 & 0.0388 & 0.0915       & 0.3400     & 0.0373     \\ \hline
\end{tabular}
\end{center}
\end{table*}

In Table \ref{tab:2}, the properties of the 21 EOS are summarized. These include the nuclear matter parameters,  defined from the usual approximation for the energy per nucleon
\bea
 \varepsilon(\rho,\delta)\simeq \varepsilon(\rho,0)+S(\rho)\delta^2,
 \label{eq:eden}
\eea 
where $\varepsilon(\rho,0)$ is the  symmetric nuclear matter (SNM) EOS,  $S(\rho)$ is the the symmetry energy   and  $\delta=(\rho_p-\rho_n)/(\rho_p+\rho_n)$ is the asymmetry.
The symmetric nuclear matter parameters defined at saturation are  the energy per nucleon $\varepsilon_0=\varepsilon(\rho_0,0)$ ($n=0$), the incompressibility coefficient $K_0$ ($n=2$), the skewness  $Q_0$ ($n=3$),  and  the kurtosis $Z_0$ ($n=4$), respectively, given by
\begin{equation}
X_0^{(n)}=3^n \rho_0^n \left (\frac{\partial^n \varepsilon(\rho, 0)}{\partial \rho^n}\right)_{\rho_0}, \, n=2,3,4;
\label{x0}
\end{equation}
The symmetry energy parameters at saturation are:  the symmetry energy at saturation 
 $J_{\rm sym,0}$ ($n=0$), 
\begin{equation}
J_{\rm sym,0}= S(\rho_0)=\frac{1}{2} \left (\frac{\partial^2 \varepsilon(\rho,\delta)}{\partial\delta^2}\right)_{\delta=0},
\end{equation}
the slope $L_{\rm sym,0}$ ($n=1$),  the curvature $K_{\rm sym,0}$ ($n=2$),  the skewness $Q_{\rm sym,0}$ ($n=3$), and  the kurtosis $Z_{\rm sym,0}$ ($n=4$), 
respectively, defined as
\begin{equation}
X_{\rm sym,0}^{(n)}=3^n \rho_0^n \left (\frac{\partial^n S(\rho)}{\partial \rho^n}\right )_{\rho_0},\, n=1,2,3,4.
\label{xsym}
\end{equation}

Looking at Table \ref{tab:2}, the binding energy per nucleon, denoted as \(\epsilon_0\), displays a median value of -16.16 MeV, with its values spanning from -16.40 MeV to -15.72 MeV. The incompressibility of nuclear matter, represented by \(K_0\), had a median of 233 MeV, and its value varies between 190 MeV and 290 MeV. The skewness of the data, indicated by \(Q_0\), had a median of -339 MeV, with an extensive range from -491 MeV to 375 MeV. Regarding the transition density, symbolized as \(\rho_t\), the median was calculated at 0.0944 fm$^{-3}$, with its range stretching from 0.067 fm$^{-3}$ to 0.127 fm$^{-3}$. Lastly, the nuclear saturation density, \(\rho_0\), shows a more consistent agreement among the EOS models, with a median of 0.155 fm$^{-3}$ and values ranging from 0.144 fm$^{-3}$ to 0.165 fm$^{-3}$.


\subsection{Equation of state}

\begin{figure}
    \centering
    \includegraphics[width=0.99\linewidth]{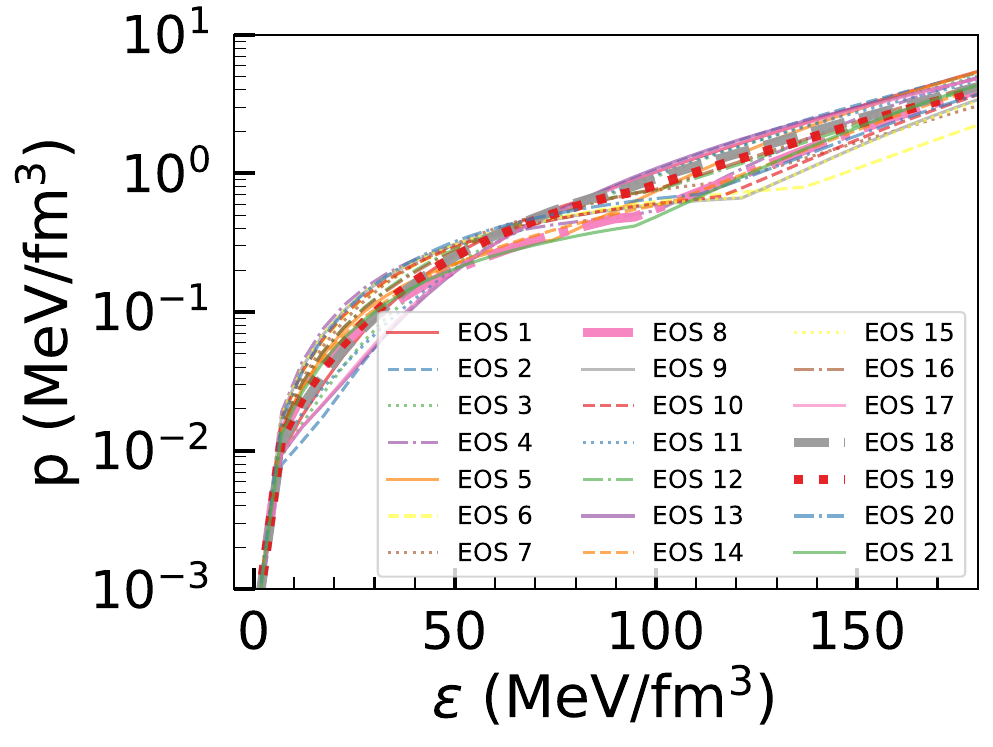}
    \caption{Pressure versus energy density for neutron star matter in $\beta$-equilibrium, using our EOS set.}
    \label{fig:eos}
\end{figure}

In Figure \ref{fig:eos}, the EoS of all the 21 models considered in this work are shown: the pressure is plotted as a function of the energy density for homogeneous stellar matter, with an inner crust calculated in the CLD method. This figure highlights the crust part, where the first branch, up to $\epsilon\sim 10$ MeV.fm$^{-3}$, represents the BSk22 outer crust, and the following branch, up to  $\epsilon\sim 100$  MeV.fm$^{-3}$, represents the inner crust, as the value $\sim 150$ MeV.fm$^{-3}$ roughly corresponds to the nuclear saturation density, 0.15 fm$^{-3}$. At the transition to the core, the EoS that has the lowest value of the pressure is EoS 5, with $P_t=0.318$ MeV.fm$^{-3}$, and the one that has the highest value is EoS 7, with $P_t=0.7956$ MeV.fm$^{-3}$. These values are written in the previous Table \ref{tab:2}.

\begin{figure}
    \centering
    \includegraphics[width=0.99\linewidth]{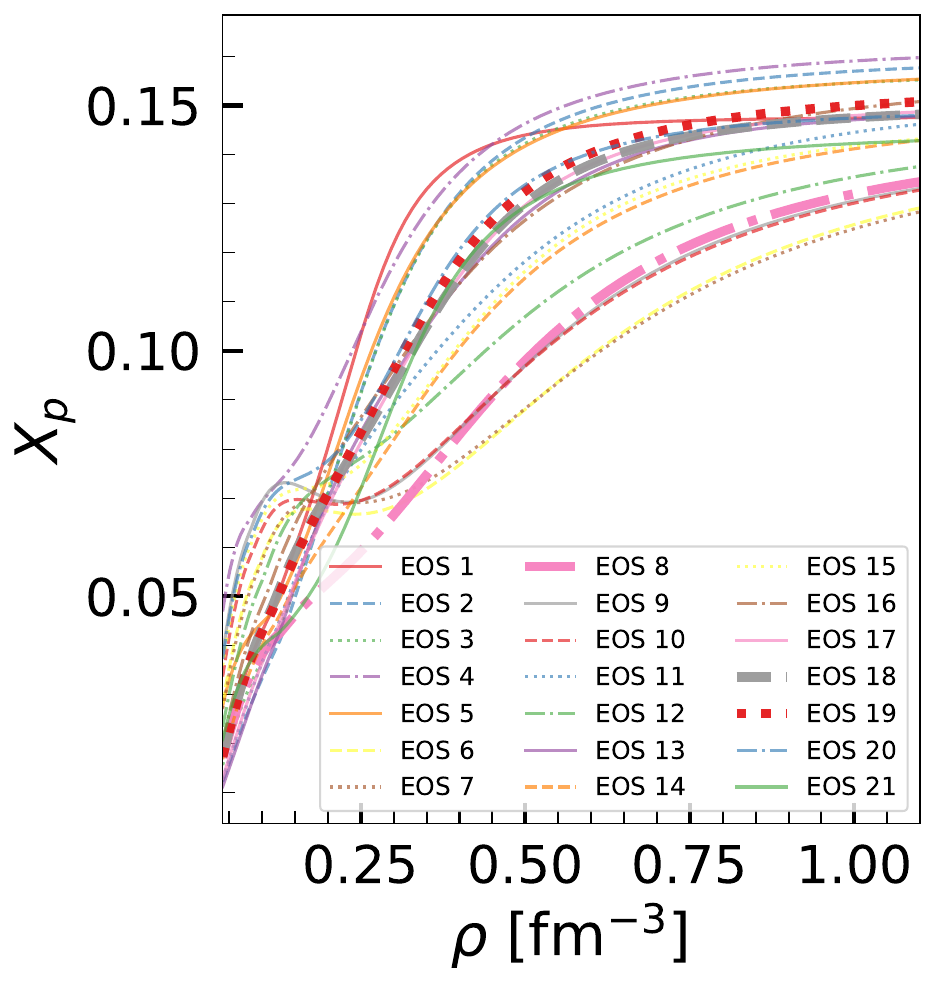}
    \caption{The proton fraction as a function baryon density for all the twenty-one EOS considered.}
    \label{fig:xp}
\end{figure}

Fig.~\ref{fig:xp} illustrates the intricate relationship between the proton fraction and density within neutron stars. This is a fundamental link to the symmetry energy of the model. The direct Urca (dUrca) threshold value of the proton fraction is particularly important.  At this density  the dUrca processes, a major factor in the cooling of NS, are triggered. The internal chemical composition of neutron stars (NS) is established through the weak interaction, specifically the \(\beta\) equilibrium condition. The composition has a major impact on the cooling efficiency of both middle-aged isolated neutron stars and accreting neutron stars. The dUrca process, which involves the weak interaction-based conversion of neutrons into protons and vice versa, is much more efficient — approximately a million times — than the modified Urca process which occurs mediated by several nucleons, as highlighted in the research conducted by Yakovlev and collaborators \cite{Yakovlev:2000jp,Yakovlev:2004iq}. The onset of the dUrca process requires a certain threshold in the proton fraction, specifically a minimum of 1/9 if no muons or other charged species are considered, as pointed out by \cite{Lattimer:1991ib}. This threshold proton fraction increases slightly after the onset of muons. The proton fraction within neutron stars, denoted as \( y_p \), increases with baryonic density. Therefore, in massive NSs, if \( y_p \) surpasses the critical threshold for the proton fraction, dUrca processes can potentially start.
Recent studies, such as those by \cite{Beznogov:2015ewa}, have used statistical approaches to describe the thermal evolution of isolated and accreting NSs. These studies suggest that a successful description of cooling curves is achievable by considering the onset of the dUrca processes in stars with masses of the order of 1.6 to 1.8 solar masses. {Table \ref{tab:3} provides the dUrca threshold density $\rho_{\rm dUrca}$ and the corresponding NS mass Urca mass where the dUrca sets in at the center, $M_{\rm dUrca}$, for all models. Some models do no predict the occurrence of nucleonic dUrca in the NS interior. In this case, a slower, but still quite efficient process may occur if hyperons set in. Several models present the onset of the nucleonic dUrca inside stars with masses $\sim 1.6$ to $1.8\, M_\odot$, but for most of the NS which allow nucleonic dUrca, it only occurs inside stars with a mass of the order of 2 $M_\odot$ or larger. Notice, however, that if hyperonic degrees of freedom are allowed, hyperonic dUrca processes may occur inside NS with smaller masses, since the hyperonic dUrca processes open for densities close to the onset hyperon density.}

\begin{table}[]
\centering
\setlength{\tabcolsep}{18.pt}
\renewcommand{\arraystretch}{1.2}
\caption{All the employed RMF EOS are converted to Spectral representation EOS format \cite{Lindblom:2010bb} in pressure based forms, where: $\epsilon_0/c^2=2.03 \times 10^{14}$ g/cm$^3$ and $x_{max}=10$. These are the parameters that have been fitted, along with the highest relative error, referred to as RE \label{tab:6}. The unit of $p_0$ is dyne $\times$ cm$^{-2}$.}
\setlength{\tabcolsep}{3.0pt}
      \renewcommand{\arraystretch}{1.1}
\begin{center}
\begin{tabular}{lrrrrrr}
\hline 
\hline 
EOS & $\gamma_0$ & $\gamma_1$ & $\gamma_2$ & $\gamma_3$ & $p_0$ & \% RE \\
\hline 
EOS1 & 0.9368 & 0.5084 & -0.1512 & 0.0100 & 2.3e+33 & 0.5023 \\
EOS2 & 0.6425 & 0.7209 & -0.2287 & 0.0172 & 2.4e+33 & 0.0518 \\
EOS3 & 0.6488 & 0.7571 & -0.2360 & 0.0175 & 2.2e+33 & 0.0721 \\
EOS4 & 1.0026 & 0.4939 & -0.1660 & 0.0119 & 1.4e+33 & 0.0439 \\
EOS5 & 0.8759 & 0.5776 & -0.1959 & 0.0147 & 1.9e+33 & 0.0542 \\
EOS6 & 1.1172 & 0.1133 & -0.0317 & 0.0012 & 1.1e+33 & 0.0799 \\
EOS7 & 1.1334 & 0.0755 & -0.0249 & 0.0009 & 1.4e+33 & 0.0634 \\
{\bf EOS8} & 1.2349 & 0.0871 & -0.0326 & 0.0013 & 1.3e+33 & 0.1359 \\
EOS9 & 1.3365 & 0.0650 & -0.0342 & 0.0021 & 1.0e+33 & 0.2074 \\
EOS10 & 1.4053 & -0.0054 & -0.0168 & 0.0007 & 1.1e+33 & 0.0948 \\
EOS11 & 0.7885 & 0.3558 & -0.0918 & 0.0048 & 2.2e+33 & 0.2113 \\
EOS12 & 1.1019 & 0.1633 & -0.0494 & 0.0023 & 1.6e+33 & 0.0668 \\
EOS13 & 0.6199 & 0.5032 & -0.1221 & 0.0071 & 2.5e+33 & 0.2778 \\
EOS14 & 1.0968 & 0.2814 & -0.0868 & 0.0049 & 1.3e+33 & 0.1851 \\
EOS15 & 1.1175 & 0.2604 & -0.0715 & 0.0034 & 1.0e+33 & 0.3341 \\
EOS16 & 0.9500 & 0.3657 & -0.1142 & 0.0073 & 1.7e+33 & 0.1548 \\
EOS17 & 0.6207 & 0.5822 & -0.1499 & 0.0091 & 2.3e+33 & 0.2642 \\
{\bf EOS18} & 0.6025 & 0.6341 & -0.1615 & 0.0098 & 2.0e+33 & 0.2765 \\
{\bf EOS19} & 0.6244 & 0.6703 & -0.1802 & 0.0115 & 1.8e+33 & 0.2340 \\
EOS20 & 0.7339 & 0.6448 & -0.1634 & 0.0097 & 1.2e+33 & 0.2055 \\
EOS21 & 0.9125 & 0.5346 & -0.1416 & 0.0088 & 1.2e+33 & 0.4828 \\
\hline
\end{tabular}
\end{center}
\end{table}

Astrophysicists often opt for the functional form of the equation of state (EOS) instead of using tabulated data for computationally expensive calculations \cite{LIGOScientific:2018cki,Miller:2019cac,Raaijmakers:2019dks}. We also present our 21 set of EOS in the spectral representation of realistic EOS, which offers a precise and efficient method to approximate any realistic EOS with an error of less than 0.5\% \cite{Lindblom:2010bb}. In Table \ref{tab:6}, we show the parameters obtained from a spectral decomposition fit to the RMF EoS.

\subsection{Constraints from the low-density regime} \label{constraints}

In this section, we analyse the symmetric nuclear matter and neutron matter properties of the 21 EoS, and compare with constraints available in the literature. 

\begin{figure}
    \centering
    \includegraphics[width=0.99\linewidth]{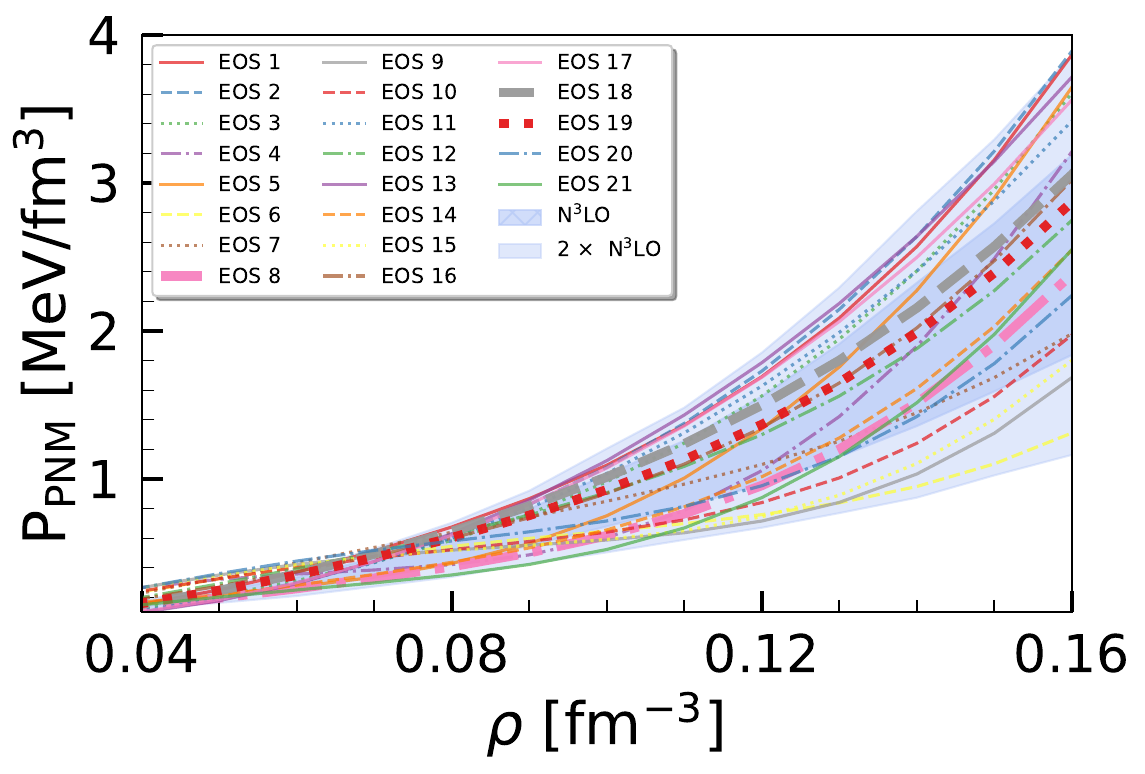}
    \caption{Pure neutron matter pressure as a function of the baryonic density. The dark (light) blue band denotes the pure neutron matter calculated within a $\chi$EFT description \cite{Hebeler2013} considering the uncertainty (twice the uncertainty) indicated.}
    \label{fig:pnm}
\end{figure}

Figure \ref{fig:pnm} displays the pressure of pure neutron matter as a function of the baryon density for the 21 EOS we analyzed. We have included the theoretical constraints derived from state-of-the-art 
N3LO calculation within a $\chi$EFT (dark grey), and the light grey represents the region where the uncertainty is doubled. All the equations of state  in our dataset are affected by N3LO uncertainty. 

\begin{figure}
    \centering
    \includegraphics[width=0.99\linewidth]{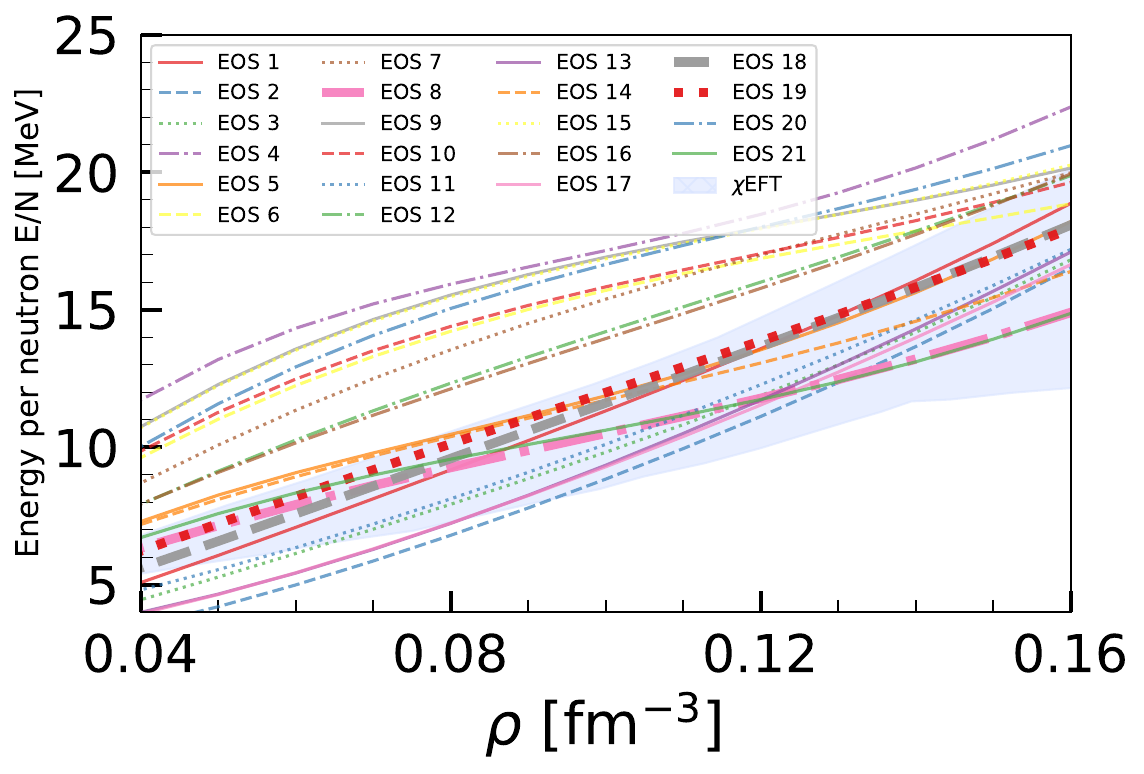}
    \caption{Energy per neutron as a function of the baryonic density. The light blue band represents enclosed region of several $\chi$EFT calculations taken from \cite{Huth22}}
    \label{fig:pnme}
\end{figure}

If instead of considering the pure neutron matter pressure given in \cite{Hebeler2013}, the pure neutron energy per particle obtained in \cite{Huth22} is imposed with the given uncertainties, several EoS are completely out of the calculated range. This is shown  in Figure \ref{fig:pnme}, where the energy per neutron is plotted as a function of density, and compared with a band that corresponds to several $\chi$EFT calculations: five EoS (EoS 1,8, 18, 19 and 21) are inside the total density range {0.04-0.16} fm$^{-3}$, and 11 EoS are inside the  range 0.08-0.16 fm$^{-3}$ (see Table \ref{tab:5} at the end for a complete list of all satisfied constraints).

\begin{figure}
    \centering
    \includegraphics[width=0.99\linewidth]{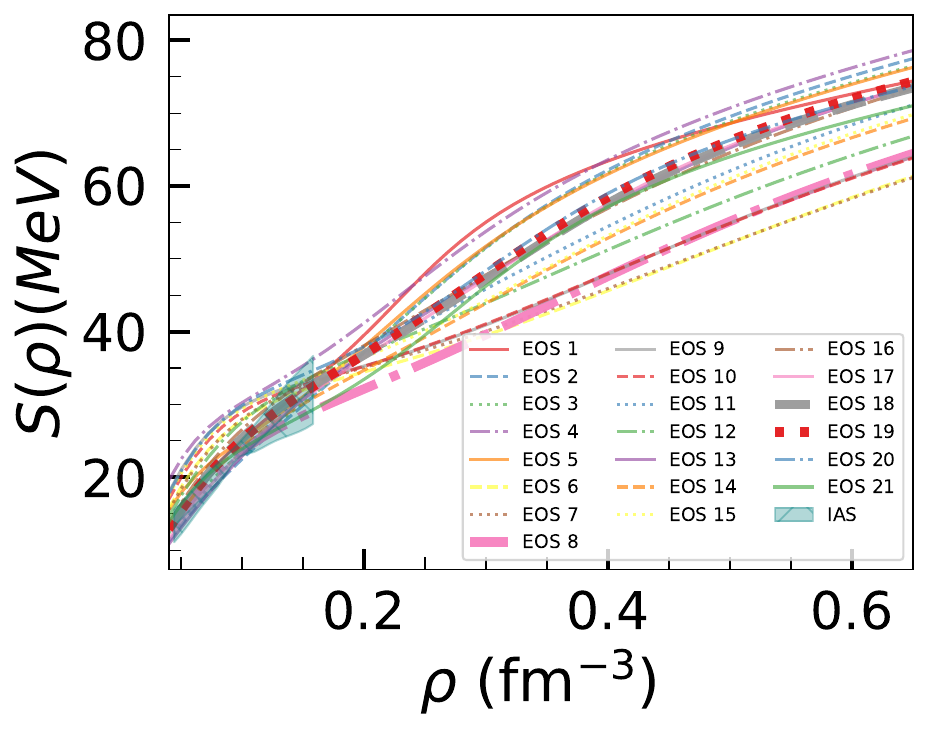}
    \caption{The symmetry energy $S(\rho)$ as a function of density $\rho$ for all the EOS. The constraints on the symmetry energy from IAS  \cite{Danielewicz:2013upa} is also displayed.}
    \label{fig:sym}
\end{figure}

In Figure \ref{fig:sym}, the symmetry energy is shown and compared with the band extracted in \cite{Danielewicz:2013upa} from  isobaric analog states (IAS). Twelve EoS satisfy this constraint, see Table \ref{tab:5} for a complete list.

\subsection{Neutron star properties }

\begin{figure}
    \centering
    \includegraphics[width=0.99\linewidth]{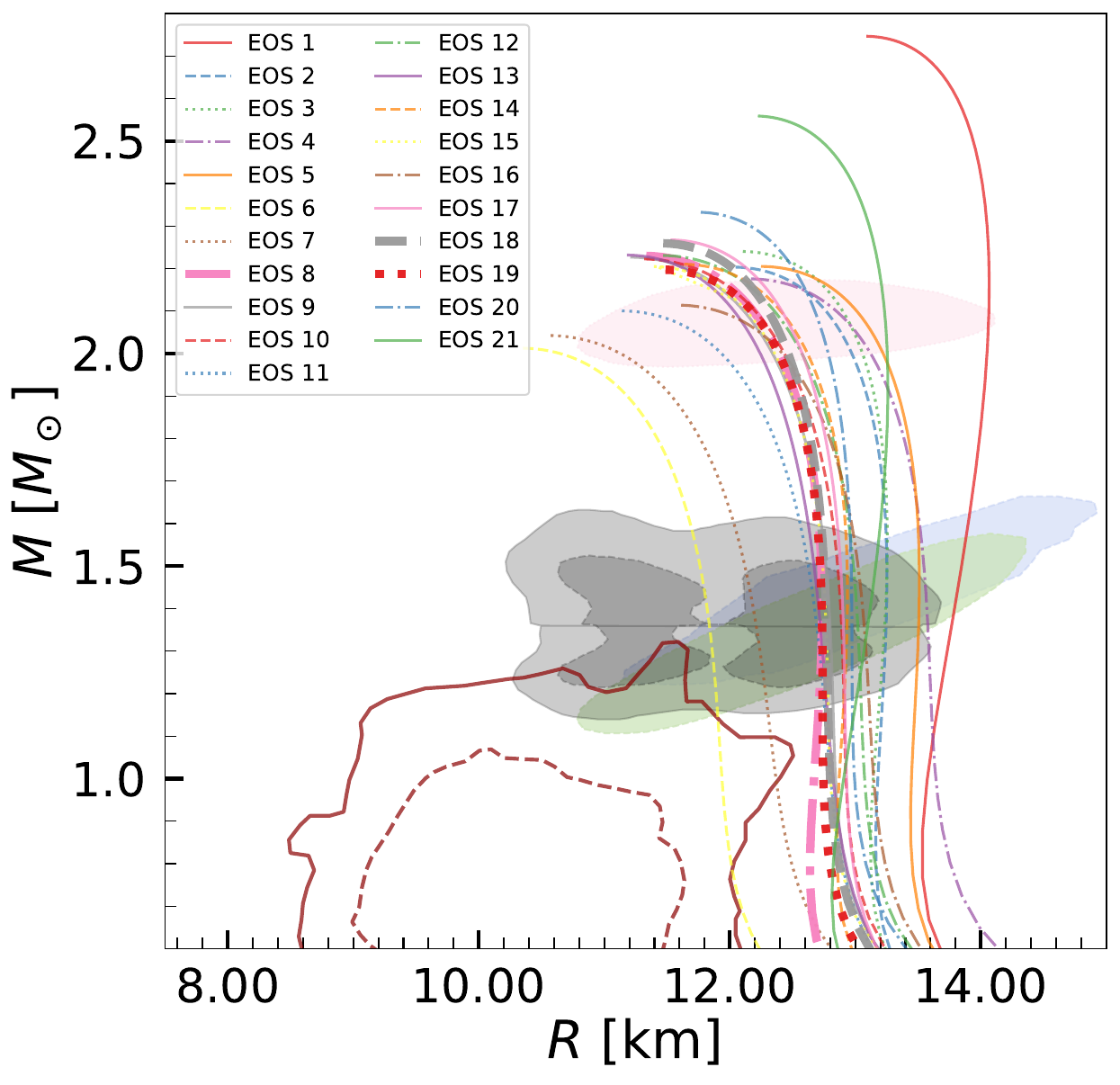}
    \caption{NS mass-radius curves for all the twenty-one models with unified crust. The gray regions indicate the 90\% (light) and 50\% (dark) credible  intervals (CI) constraints from the binary components of GW170817.  The $1\sigma$ (68\%) CI for the 2D posterior distribution in the mass-radii domain for the millisecond pulsar PSR J0030 + 0451 (cyan and green) 
\cite{Riley:2019yda, Miller:2019cac} as well as the PSR J0740 + 6620 (pink) \cite{Riley:2021pdl, Miller:2021qha}  from the NICER x-ray data are also shown. Additionally, we show the constraint obtained from HESS J1731-347 for 68.3\% (95.4\%) CIs in dashed dark red (solid dark red) \cite{hess}. }
    \label{fig:mr}
\end{figure}

In Figure \ref{fig:mr}, we show the mass-radius relations for all the models considered. In this figure several observational constraints are reported, as identified in the caption of Fig. \ref{fig:mr}. In particular, we conclude that all EoS satisfy the constraints from the gravitational wave event GW170817 at 90\% CI, except EoS1. Also almost all EoS satisfy the constraints coming from NICER both for the PSR J0030 + 0451 and  PSR J0740 + 6620 at 68\% CI: only EoS6 does not fall inside the 1$\sigma$ distribution obtained for PSR J0740 + 6620. We have also included the 1 and 2$\sigma$ 2D distributions for the low mass compact object J1731-347 recently detected by the HESS (High Energy Stereoscopic System) collaboration. From all the EoS only EoS6 and EoS19 fall inside the 2$\sigma$ CI.

\begin{figure}
    \centering
    \includegraphics[width=0.99\linewidth]{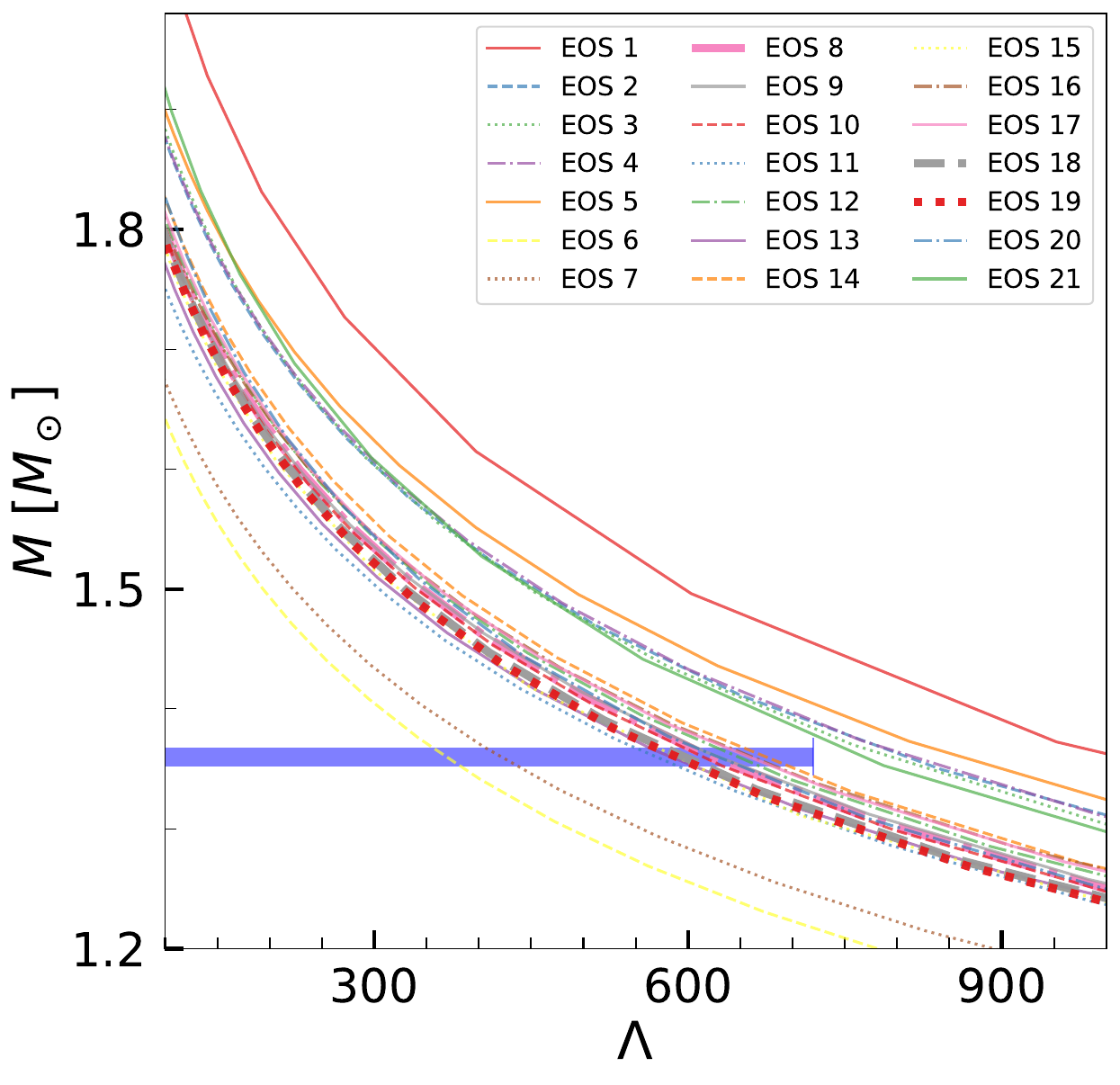}
    \caption{The mass-tidal deformability relationship of neutron stars has been determined by taking into account all equations of state with a unified crust. The blue band indicates the tidal deformability for a mass ratio of $q=1$ or for a neutron star of 1.36$M_\odot$ that was part of the binary neutron star event GW170817 \cite{LIGOScientific:2018cki}.
    }
    \label{fig:lam}
\end{figure}

\begin{figure}
    \centering
    \includegraphics[width=0.99\linewidth]{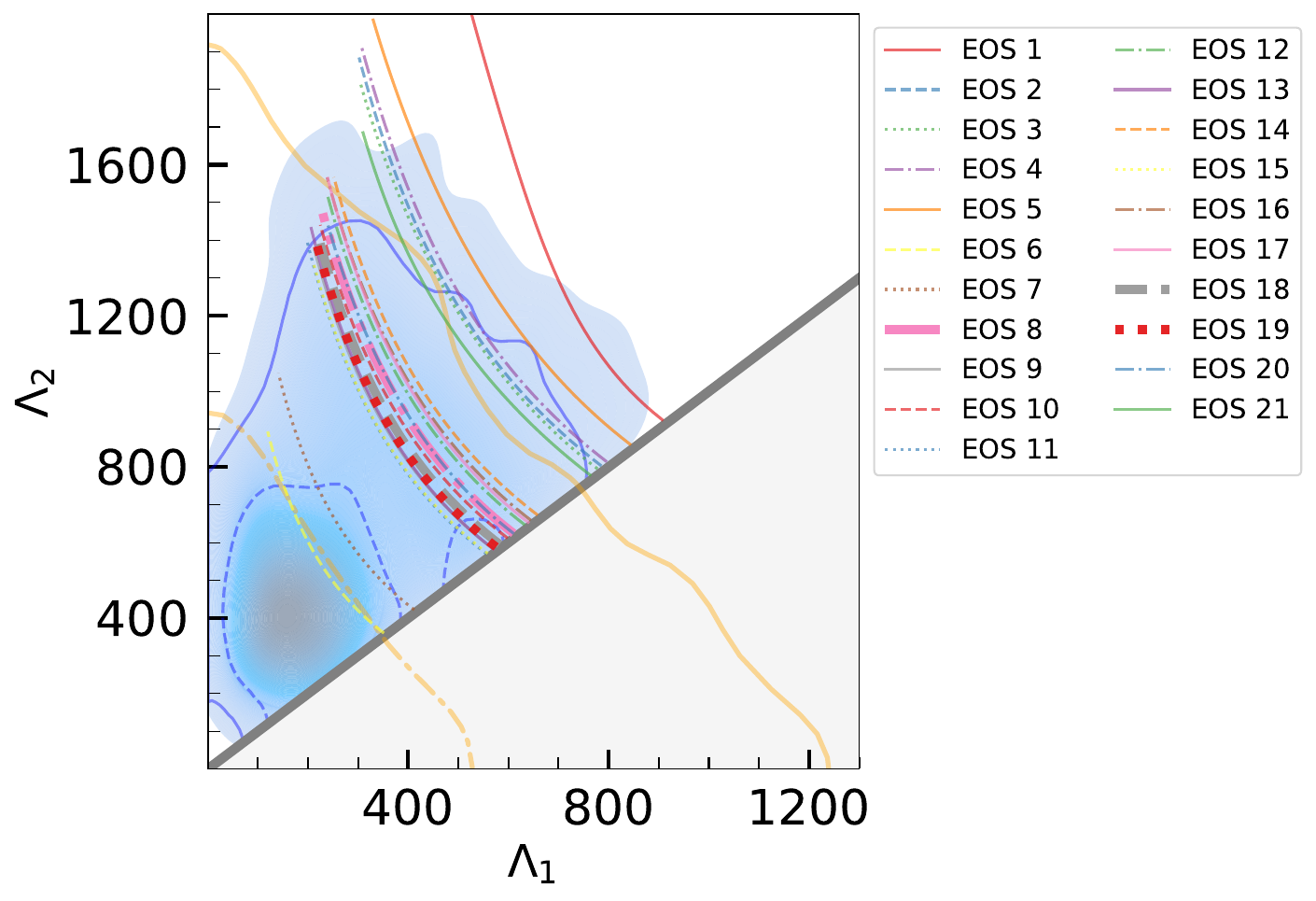}
    \caption{The dimensionless tidal deformability parameters, denoted as \( \Lambda_1 \) and \( \Lambda_2 \), are derived from the binary neutron star merger of the GW170817 event. These values are evaluated based on all 21 RMF EoSs, taking into account the observed chirp mass \( M_{\rm chirp}=1.186 \) M\(_\odot\). The orange line (solid for 90\% confidence and dashed for 50\% confidence) indicates the confidence intervals, while the blue shaded area highlights the marginalized posterior derived from a parameterized EoS. This EoS has a maximum mass stipulation of 1.97 M\(_\odot\), and within this blue region, the solid and dashed lines again depict 90\% and 50\% confidence intervals, respectively.}
    \label{fig:lam1_lam2}
\end{figure}

The GW170817 event \citep{LIGOScientific:2018cki} sets also constraints on the tidal deformability of stars with a mass of the order of 1.4$M_\odot$. In Figs. \ref{fig:lam} and \ref{fig:lam1_lam2},  the combined tidal deformability and dimensionless tidal deformabilities are shown and compared with the values from the GW170817 event. The blue band in Fig. \ref{fig:lam} identifies the constraint deduced from the tidal deformability of a 1.36$M_\odot$ NS corresponding to the  mass ratio of $q=1$ between the masses of the binary NS, which within 90\% CI is below 720. We conclude that only six EoS do not satisfy this constraint (EoS 1, 2, 3, 4, 5 and 21).
In Fig. \ref{fig:lam1_lam2}, the dimensionless tidal deformability parameters of both stars of the binary neutron star merger of the GW170817 event, \( \Lambda_1 \) and \( \Lambda_2 \), are plotted for all the 21 RMF EoSs, taking into account the observed chirp mass \( M_{\rm chirp}=1.186 \) M\(_\odot\), and compared with the data from the GW170817 event.  The orange line (solid for 90\% confidence and dashed for 50\% confidence) indicates the confidence intervals obtained in \cite{LIGOScientific:2017ync}, while the blue shaded area highlights the marginalized posterior derived from a parameterized EoS imposing a maximum mass stipulation of 1.97 M\(_\odot\) by the LVC \cite{LIGOScientific:2018hze}. Within the blue region, the solid and dashed lines  depict 90\% and 50\% confidence intervals, respectively  (see \cite{LIGOScientific:2018hze}). Considering the marginalized posterior at 90\% CI only two EoS do not satisfy the constraints, EoS 1 and 5. 

In Table \ref{tab:3}, we summarize the NS properties of the 21 EoS, in particular we give  the maximum mass $M_{max}$ , the radius of the maximum mass star $R_{max}$, the radius for a 1.4 solar mass neutron star $R_{1.4}$, the radius for a 2.08 solar mass neutron star $R_{2.08}$, the dimensionless tidal deformability for a 1.4 solar mass neutron star $\Lambda_{1.4}$, the square of the speed of sound $c_s^2$, the central baryonic density $\rho_c$, the dUrca onset density $\rho_{\rm dUrca}$, and the NS mass where nucleonic dUrca processes open, $M_{\rm dUrca}$.
We conclude that except for EoS 1 and 21, all the EoS predict a central baryonic density $\rho_c\gtrsim 6\rho_0$. Concerning the speed of sound, five EoS predict $c_s^2<0.5$, ten EoS have $c_s^2>0.6$, but never above 0.8.

\begin{table*}
\centering
\caption{For all twenty-one EOS models, we list several properties of neutron stars, such as the maximum mass $M_{\rm max}$, maximum radius $R_{\rm max}$, radius for a 1.4 solar mass neutron star $R_{1.4}$, radius for a 2.08 solar mass neutron star $R_{2.08}$, the dimensionless tidal deformability for a 1.4 solar mass neutron star $\Lambda_{1.4}$, the square of the speed of sound $c_s^2$, the central baryonic density $\rho_c$, the durca and $M_{\rm durca}$. \label{tab:3}}  
\setlength{\tabcolsep}{14.0pt}
\renewcommand{\arraystretch}{1.1}
\begin{tabular}{cccccccccc}
\hline
\hline
   EOS &  $M_{\rm max}$ &  $R_{\rm max}$ &    $R_{1.4}$ &   $R_{2.08}$ &   $\Lambda_{1.4}$ &  $c_s^2$ &  $\rho_c$ & $\rho_{\rm dUrca}$ & $M_{\rm dUrca}$ \\
     &  $[M_{\odot}]$ &  [km] &    [km] &   [km] &    &  $[c^2]$ &  [fm$^{-3}$] & [fm$^{-3}$] & $[M_{\odot}]$ \\
\hline
  EOS1 &   2.74 &  13.03 & 13.78 & 14.04 &    844 &   0.713 &  0.683 & 0.366 & 2.06 \\
  EOS2 &   2.20 &  11.97 & 13.18 & 12.79 &    658 &   0.435 &  0.877 & 0.432 & 1.74 \\
  EOS3 &   2.24 &  12.06 & 13.19 & 12.91 &    654 &   0.444 &  0.856 & 0.437 & 1.80 \\
  EOS4 &   2.17 &  12.05 & 13.37 & 12.84 &    657 &   0.419 &  0.873 & 0.394 & 1.59 \\
  EOS5 &   2.20 &  12.16 & 13.36 & 13.00 &    709 &   0.414 &  0.849 & 0.443 & 1.83 \\
  EOS6 &   2.01 &  10.28 & 11.73 &  0.00 &    291 &   0.665 &  1.168 & ... & ... \\
  EOS7 &   2.04 &  10.51 & 12.14 &  0.00 &    341 &   0.675 &  1.129 & ... & ... \\
  {\bf EOS8} &   2.23 &  11.23 & 12.57 & 12.17 &    511 &   0.658 &  0.955 & ... & ... \\
  EOS9 &   2.23 &  11.13 & 12.67 & 12.12 &    473 &   0.731 &  0.975 & ... & ... \\
  EOS10 &  2.22 &  11.23 & 12.76 & 12.21 &    499 &   0.684 &  0.969 & ... & ... \\
  EOS11 &  2.10 &  11.08 & 12.55 & 11.53 &    462 &   0.543 &  1.025 & ... & ... \\
  EOS12 &  2.23 &  11.39 & 12.92 & 12.37 &    529 &   0.634 &  0.953 & 0.829 & 2.07 \\
  EOS13 &  2.23 &  11.11 & 12.60 & 12.04 &    474 &   0.689 &  0.982 & ... & ... \\
  EOS14 &  2.21 &  11.53 & 12.77 & 12.38 &    551 &   0.527 &  0.926 & 0.644 & 2.05 \\
  EOS15 &  2.20 &  11.31 & 12.66 & 12.15 &    467 &   0.577 &  0.956 & 0.981 & ... \\
  EOS16 &  2.11 &  11.53 & 12.96 & 12.07 &    538 &   0.466 &  0.949 & 0.938 & 2.20 \\
  EOS17 &  2.27 &  11.45 & 12.78 & 12.40 &    534 &   0.603 &  0.929 & 0.660 & 2.01 \\
  {\bf EOS18} &  2.26 &  11.40 & 12.70 & 12.31 &    487 &   0.596 &  0.939 & 0.611 & 2.09 \\
  {\bf EOS19} &  2.20 &  11.41 & 12.65 & 12.20 &    481 &   0.519 &  0.949 & 0.606 & 2.07 \\
  EOS20 &  2.33 &  11.70 & 12.91 & 12.66 &    514 &   0.597 &  0.885 & 0.562 & 1.94 \\
  EOS21 &  2.56 &  12.13 & 12.95 & 13.14 &    638 &   0.767 &  0.791 & 0.551 & 2.08 \\
\hline
\end{tabular}
\end{table*}

\subsection{Crust-core transition properties and correlations}

In the present subsection we discuss some crust-core transition properties.  We have considered two different calculations to determine the transition density: the CLD \citep{Pais15} and the dynamical spinodal \citep{Pais:2016xiu} calculation. In the first one we determine the non-homogeneous matter inside the inner crust considering for the  heavy clusters a CLD description within a Wigner-Seitz approximation. In the second approach the transition density is defined by the zero eigenmode obtained when density fluctuations in infinite nuclear matter in $\beta$-equilibrium are considered. In Table \ref{tab:2}, we show along with the saturation properties, the density, and correspondent pressure and proton fraction, at the crust-core transition, respectively $\rho_t,\, P_t, \, y_{p,t}$. 
 In principle, if the calculation of the non-homogeneous matter had been carried out in a self-consistent calculation as the Thomas Fermi description in \cite{Avancini:2010ch}, we would expect the prediction from the dynamical spinodal to be slightly below the CLD one. However,  the CLD calculation is not totally self-consistent due to the surface tension considered. There are six EoS for which the spinodal transition is above the CLD but the difference is $\lesssim 2\%$. The values predicted are of the order of $\rho_0/2- 2\rho_0/3,$ the smallest being $\sim 0.07$fm$^{-3}$ and the largest $\sim 0.13$fm$^{-3}$. At the transition, the pressure varies between $\sim 0.35$ and 0.70 MeV/fm$^3$. Another quantity that is also interesting is the proton fraction at the crust core transition. For the 21 EoS, it is generally above 0.03 (except for 3 EoS) and can be as high as 0.065. This is a quantity that is defining the behavior of the crust in the presence of a strong magnetic field as discussed in \cite{Fang:2016kcm,Fang:2017zsb,Wang:2022sxx}: a low value of $y_p$ may imply a larger inner crust.

\begin{figure}
    \centering
    \includegraphics[width=0.99\linewidth]{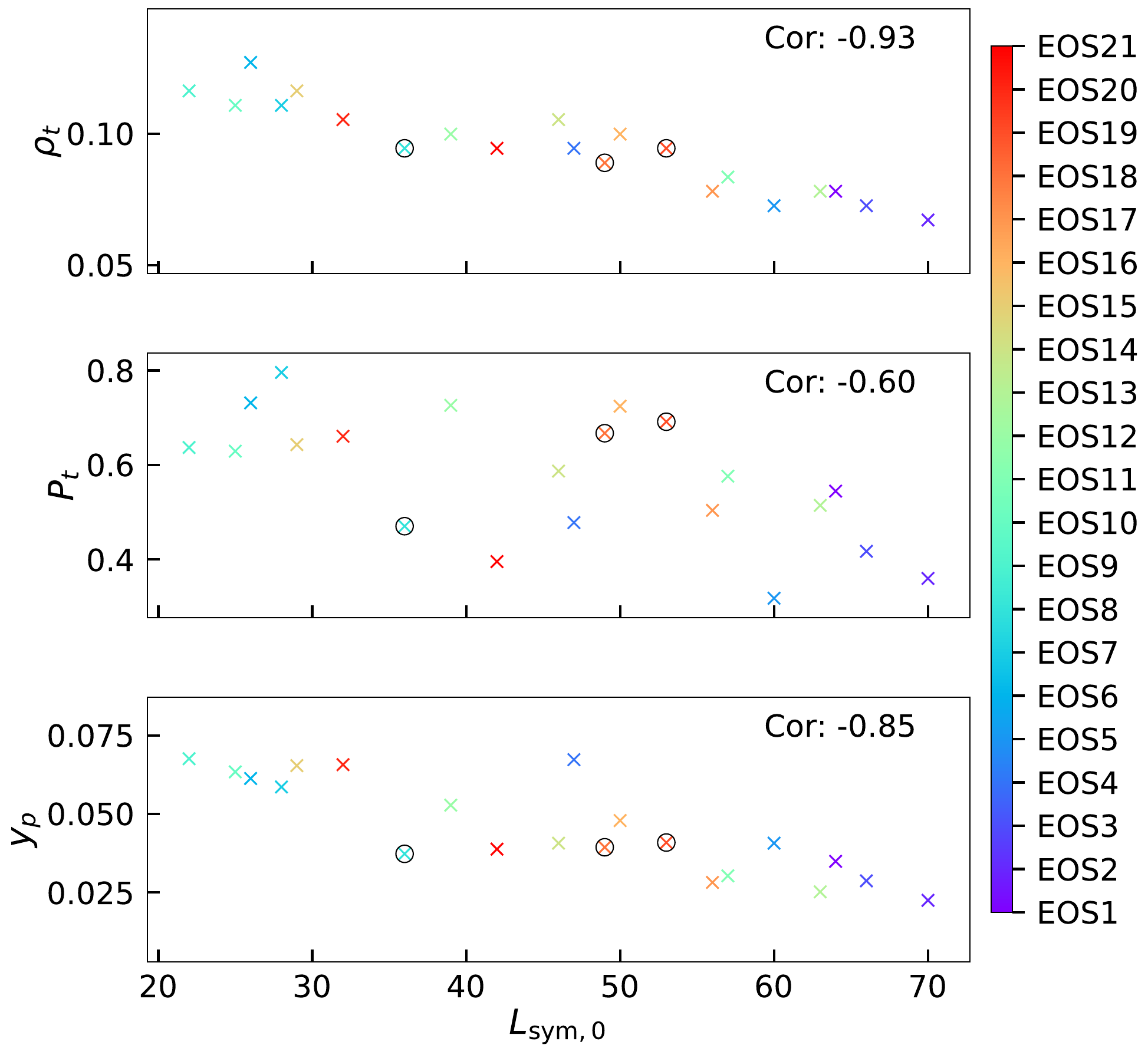}
    \caption{From top to bottom, the figure illustrates the dependencies of the transition density ($\rho_t$), $\beta$-equilibrium pressure ($p_t$), and proton fraction ($y_p$) at core crust transition density over the symmetry energy slope parameter ($L_{\rm sym,0}$).}
    \label{fig:rhot}
\end{figure}

In \cite{Ducoin:2010as,Ducoin:2011fy}, it was shown that the crust-core transition density is well correlated with the slope of the symmetry energy at saturation $L_{\rm sym,0}$. Since the 21 EoS have been precisely chosen by the value of this parameter, the 21 EoS span a quite large range, $22\le L_{\rm sym,0} \le 70$ MeV. In Figure \ref{fig:rhot}, we plot the transition density, pressure and proton fraction, respectively, $\rho_t$, $P_t$ and $y_{p,t}$ in terms of the symmetry energy slope $L_{\rm sym,0}$. As discussed in \cite{Ducoin:2010as,Ducoin:2011fy}, the transition density $\rho_t$ is very well correlated with $L_{\rm sym,0}$. Also the proton fraction $y_{p,t}$ shows a quite  good correlation with $L_{\rm sym,0}$, but not the pressure $P_t$.

\begin{table*}[]
\setlength{\tabcolsep}{10.pt}
\renewcommand{\arraystretch}{1.2}
\caption{Relations among core-crust transitional properties, i.e. the transition density \( \rho_t \), and the \( \beta \)-equilibrium pressure \( P_t \), and proton fraction \( y_{p,t} \) at the transition density, with nuclear saturation properties. The absolute Pearson correlation coefficient $r_{x,y}$ for each relationship is also listed. Nuclear saturation properties are evaluated in MeV at saturation density $\rho_0$ fm$^{-3}$. The units of $\rho_t$ and $P_t$ are fm$^{-3}$ and MeV.fm$^{-3}$, respectively. The proton fraction $y_{p,t}$ calculated at the transition density $\rho_t$ is a dimensionless quantity. We also provide the average relative percentage of errors for each relation.}
\label{tab:4}
\begin{tabular}{clcc}
\hline \hline 
Quantity                  & Relations                                                                                                                                                                                 & abs($r_{x,y}$) & \% Avg. RE \\ \hline
\multirow{3}{*}{$\rho_t$} & $\rho_t$= -0.001 $\times$ $L_{\rm sym,0}$ + 0.1417                                                                                                                                        & 0.93           & 4      \\
                          & $\rho_t$= $\rho_0$ (0.955 - 0.0075 $\times$ $L_{\rm sym,0}$)                                                                                                                               & 0.97           & 2.5       \\
                          & $\rho_t$=$\rho_0$ - 0.042 $\times$ $L_{\rm sym,0}$/$J_{\rm sym,0}$                                                                                                                        & 0.98           &  2.7       \\
                          &                                                                                                                                                                                           &                &         \\
\multirow{2}{*}{$P_t$}    & $P_t$= 0.0334 $\times$ $J_{\rm sym,0}$ - 0.00019 $\times$ $K_{\rm sym,0}$  + 0.00019 $\times$  $Q_{\rm sym,0}$ - 0.7246 & 0.90            & 7.2      \\
                          & $P_t$= $J_{\rm sym,0}$ $\times$ (-2.2433e-5 $\times$ $K_{\rm sym,0}$  + 4.5913e-6 $\times$ $Q_{\rm sym,0}$ + 0.0117)      & 0.93           & 7.9      \\
                          &                                                                                                                                                                                           &                &         \\
$y_p$                     & $y_p$ = 0.0022 $\times$ $J_{\rm sym,0}$ - 0.0005 $\times$ $L_{\rm sym,0}$                                                                                                                 & 0.95           & 11      \\ \hline
\end{tabular}
\end{table*}

In Table \ref{tab:4}, we present a fit of these correlations in terms of the symmetry energy parameters $J_{\rm sym,0}$, $L_{\rm sym,0}$ ($\rho_t$ and $y_{p,t}$). For the pressure we also consider the two higher order parameters $K_{\rm sym,0}$ and $Q_{\rm sym,0}$. The absolute Pearson correlation coefficient $r_{x,y}$ for each relationship is also listed, together with the relative mean square error (RMSE). The stronger correlations with the smallest RMSE are precisely the ones obtained for the transition density.

\begin{table*}[]
\centering
\caption{The neutron star radius for masses between 1.4 and 2.0 solar masses, using the unified crust (CLD approach, as described in section \ref{cld}), and the BPS + polytropic crust (as outlined in \cite{Malik:2023mnx}), is listed here with the corresponding percentage error. }
\label{tab:unified}
\setlength{\tabcolsep}{7.pt}
\renewcommand{\arraystretch}{1.2}
\begin{tabular}{ccccccccccccc}
\hline \hline 
\multirow{2}{*}{Model} & \multicolumn{4}{c}{With unified crust}        & \multicolumn{4}{c}{with BPS + Polytropic crust} & \multicolumn{4}{c}{\% error}                                              \\ \cline{2-13} 
                       & $R_{1.4}$ & $R_{1.6}$ & $R_{1.8}$ & $R_{2.0}$ & $R_{1.4}$  & $R_{1.6}$  & $R_{1.8}$ & $R_{2.0}$ & $\Delta R_{1.4}$ & $\Delta R_{1.6}$ & $\Delta R_{1.8}$ & $\Delta R_{2.0}$ \\ \hline
EOS 1                  & 13.79     & 13.9      & 13.99     & 14.05     & 13.78      & 13.88      & 13.97     & 14.03     & 0.07             & 0.14             & 0.14             & 0.14             \\
EOS 2                  & 13.25     & 13.24     & 13.17     & 12.97     & 13.18      & 13.18      & 13.11     & 12.92     & 0.53             & 0.45             & 0.46             & 0.39             \\
EOS 3                  & 13.24     & 13.25     & 13.21     & 13.06     & 13.19      & 13.2       & 13.16     & 13.01     & 0.38             & 0.38             & 0.38             & 0.38             \\
EOS 4                  & 13.57     & 13.51     & 13.39     & 13.12     & 13.37      & 13.34      & 13.24     & 13        & 1.47             & 1.26             & 1.12             & 0.91             \\
EOS 5                  & 13.51     & 13.5      & 13.42     & 13.22     & 13.36      & 13.37      & 13.31     & 13.13     & 1.11             & 0.96             & 0.82             & 0.68             \\
EOS 6                  & 11.82     & 11.69     & 11.42     & 10.65     & 11.73      & 11.61      & 11.36     & 10.59     & 0.76             & 0.68             & 0.53             & 0.56             \\
EOS 7                  & 12.19     & 12.03     & 11.76     & 11.13     & 12.14      & 11.99      & 11.72     & 11.09     & 0.41             & 0.33             & 0.34             & 0.36             \\
{\bf EOS 8}                  & 12.74     & 12.71     & 12.61     & 12.39     & 12.57      & 12.57      & 12.5      & 12.29     & 1.33             & 1.1              & 0.87             & 0.81             \\
EOS 9                  & 12.79     & 12.73     & 12.6      & 12.34     & 12.67      & 12.63      & 12.51     & 12.26     & 0.94             & 0.79             & 0.71             & 0.65             \\
EOS 10                 & 12.88     & 12.82     & 12.69     & 12.44     & 12.76      & 12.72      & 12.6      & 12.36     & 0.93             & 0.78             & 0.71             & 0.64             \\
EOS 11                 & 12.68     & 12.56     & 12.36     & 11.93     & 12.55      & 12.45      & 12.27     & 11.86     & 1.03             & 0.88             & 0.73             & 0.59             \\
EOS 12                 & 13.01     & 12.95     & 12.82     & 12.57     & 12.92      & 12.87      & 12.75     & 12.51     & 0.69             & 0.62             & 0.55             & 0.48             \\
EOS 13                 & 12.71     & 12.63     & 12.49     & 12.24     & 12.6       & 12.53      & 12.4      & 12.17     & 0.87             & 0.79             & 0.72             & 0.57             \\
EOS 14                 & 12.94     & 12.92     & 12.83     & 12.62     & 12.77      & 12.78      & 12.71     & 12.52     & 1.31             & 1.08             & 0.94             & 0.79             \\
EOS 15                 & 12.76     & 12.71     & 12.6      & 12.36     & 12.66      & 12.62      & 12.52     & 12.29     & 0.78             & 0.71             & 0.63             & 0.57             \\
EOS 16                 & 13.06     & 12.97     & 12.8      & 12.42     & 12.96      & 12.88      & 12.72     & 12.35     & 0.77             & 0.69             & 0.63             & 0.56             \\
EOS 17                 & 12.9      & 12.86     & 12.76     & 12.58     & 12.78      & 12.76      & 12.68     & 12.5      & 0.93             & 0.78             & 0.63             & 0.64             \\
{\bf EOS 18}                 & 12.77     & 12.73     & 12.64     & 12.47     & 12.7       & 12.67      & 12.59     & 12.42     & 0.55             & 0.47             & 0.4              & 0.4              \\
{\bf EOS 19}                 & 12.73     & 12.7      & 12.6      & 12.39     & 12.65      & 12.63      & 12.54     & 12.33     & 0.63             & 0.55             & 0.48             & 0.48             \\
EOS 20                 & 12.97     & 12.96     & 12.91     & 12.78     & 12.91      & 12.9       & 12.85     & 12.73     & 0.46             & 0.46             & 0.46             & 0.39             \\
EOS 21                 & 13.12     & 13.2      & 13.25     & 13.26     & 12.95      & 13.06      & 13.13     & 13.14     & 1.3              & 1.06             & 0.91             & 0.9        \\ \hline      
\end{tabular}
\end{table*}

 Before finishing this sub-section, we would like to comment the approximation implemented in \cite{Malik:2023mnx} for the inner crust. In that work,  the BPS EoS was used for the outer crust, and for the inner crust,  a polytropic EoS was used. It was matched to the core EoS at 0.04fm$^{-3}$. Although this density is well below the crust-core transition, it was considered that the error introduced considering the outer core EoS already above this density would be small, because the energy difference between the homogeneous matter EoS and the non-homogeneous EoS is { small  in this range of densities}, see \cite{Avancini:2008zz}.
 In Table \ref{tab:unified} we list the  neutron star radius of stars with  masses between 1.4 and 2.0 solar masses for the 21 EoS, using the unified crust (CLD approach, as described in section \ref{cld}), and the BPS + polytropic crust (as outlined in \cite{Malik:2023mnx}) as well as  the corresponding percentage error. The maximum error introduced with the BPS + polytropic crust + core EoS at 0.04fm$^{-3}$ is of the order of 1\% - 1.5\%, but in most cases it is well below 1\%.

\subsection{Trace anomaly and speed of sound}

\begin{figure}
    \centering
    \includegraphics[width=0.99\linewidth]{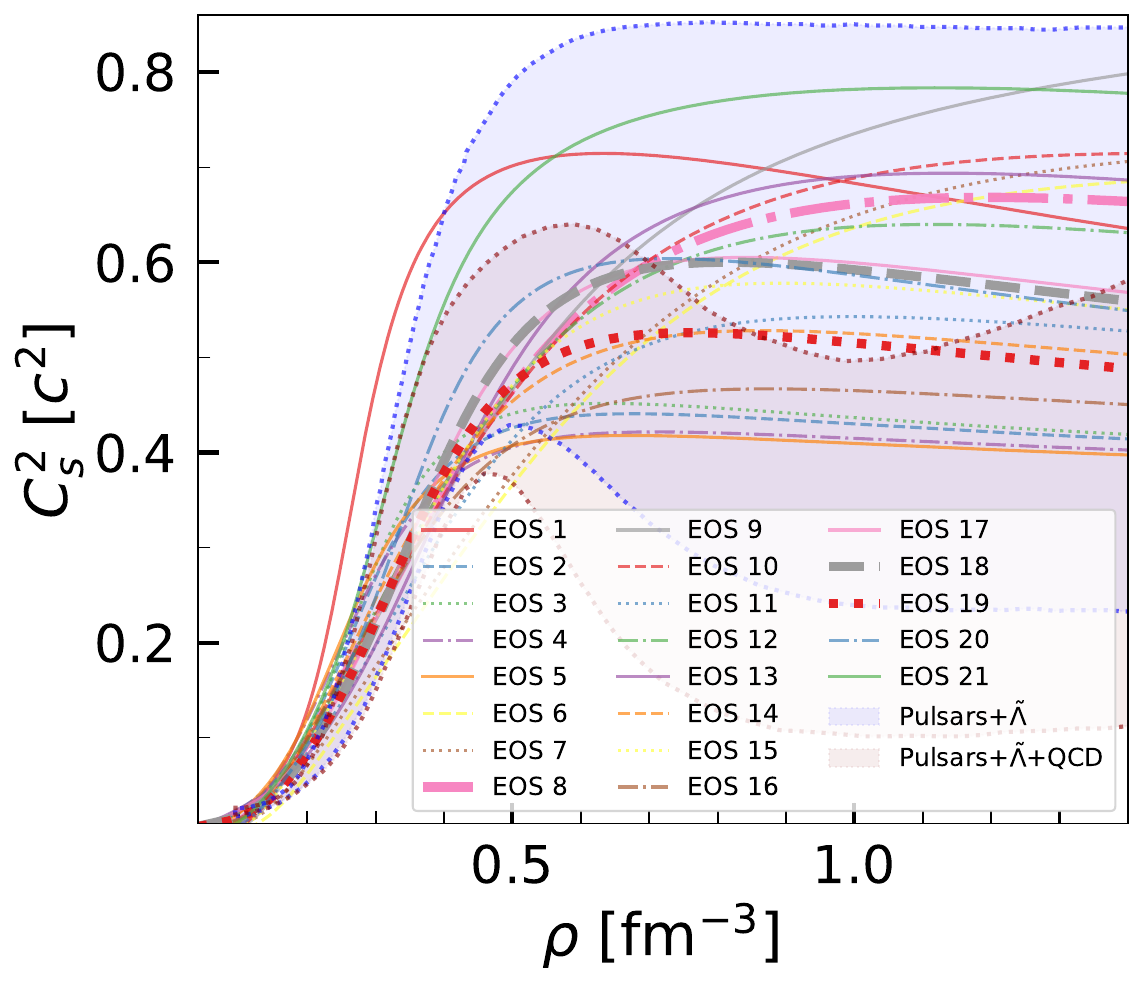}
    \caption{The square of the speed of sound $c_s^2$ as a function of density for the EOS set considered. We also compare the posteriors obtained with "astro" and "astro" + pQCD constraint in Ref.~\cite{Kurkela:2022elj}.} 
    \label{fig:cs2}
\end{figure}

\begin{figure}
    \centering
    \includegraphics[width=0.99\linewidth]{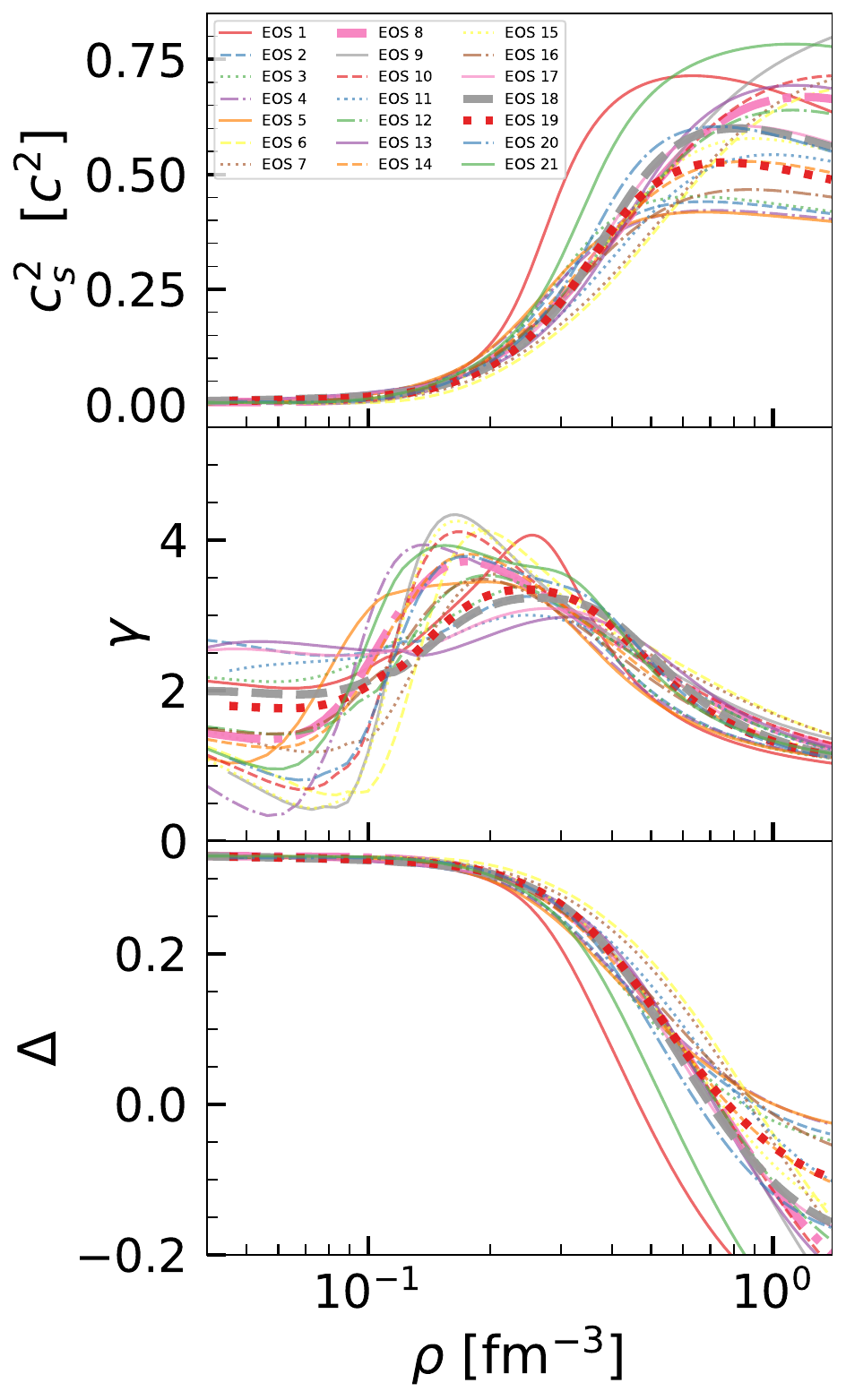}
    \caption{The speed of sound squared $c_s^2$, the polytropic index $\gamma=d \mbox{ln}P/d\mbox{ln}\epsilon$ and the trace anomaly $\Delta=1/3-P/\epsilon$ for all the EOS.}
    \label{fig:cs2x}
\end{figure}

\begin{figure}
    \centering
    \includegraphics[width=0.99\linewidth]{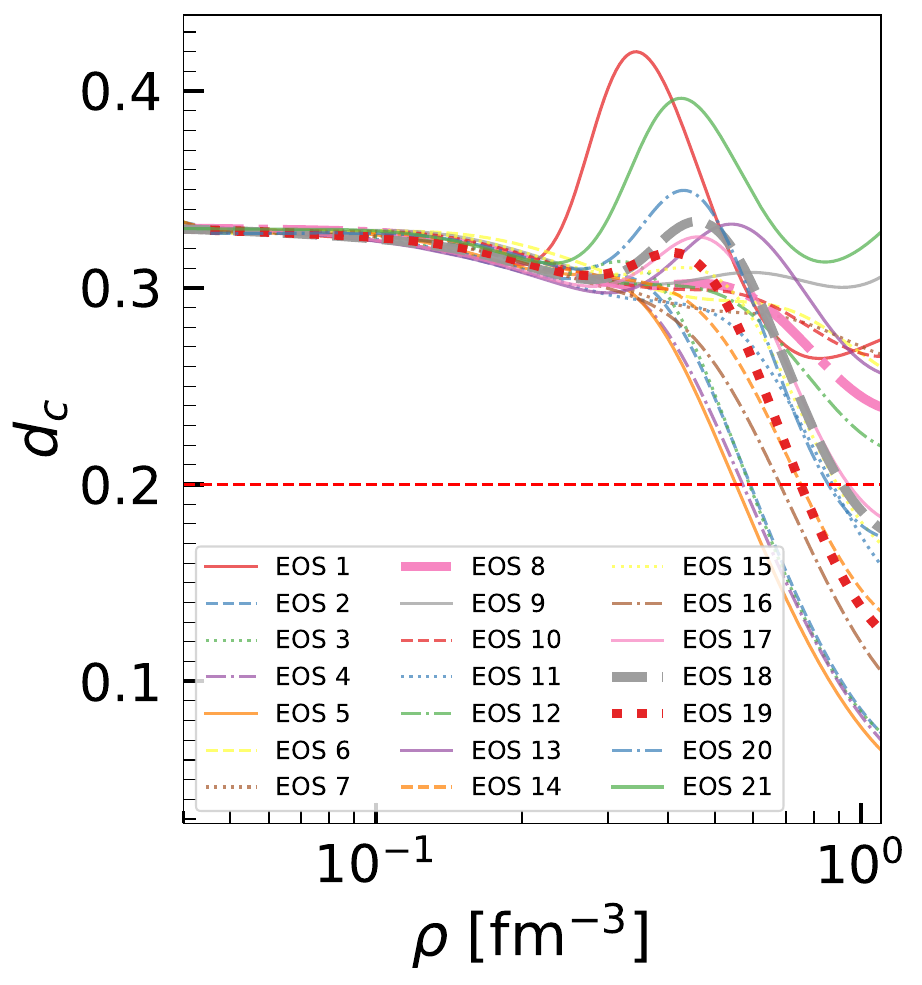}
    \caption{This figure shows the relationship between $d_c$ and $\rho$ for all EOS (see text for details).}
    \label{fig:dc}
\end{figure}
The identification of a phase transition to deconfined matter inside NS has been the topic of many recent studies developed within different frameworks, see, for instance, \cite{Annala2019,Tews:2018kmu,Altiparmak:2022bke,Gorda:2022jvk,Somasundaram:2021clp,Somasundaram:2022ztm}.  In \cite{Kurkela:2022elj}, using a Gaussian Process description of the EoS, the authors have estimated which would be the effect of the EoS obtained in the perturbative QCD  (pQCD) regime (i.e. above $\sim 40\rho_0$) on the EoS of neutron star matter. In particular, the authors have analysed the effect on the speed of sound.

In Fig. \ref{fig:cs2}, we have plotted the speed of sound squared as a function of the baryonic density for our 21 EoS. In that figure, we also include the two regions calculated in \cite{Kurkela:2022elj} which define the 90\% CI for the square of the speed of sound  imposing only astrophysical constraints (the blue band denominated Astro) and astrophysical together with  pQCD constraints (the brown band denominated Astro + pQCD). One of the consequences of imposing the pQCD constraints is to push the speed of sound to lower values at high densities and, in particular, to give raise to a peak at a baryonic density of the order of $\sim 0.5$ fm$^{-3}$.  The 21 EoS presented in our study are nucleonic EoS, without any kind of phase transition. No EoS shows a peak on the speed of sound at $0.5$ fm$^{-3}$, but a large number flattens above this density, and some even stay inside the Astro + pQCD band for the whole range of densities, which may imply that the identification of the presence of a quark phase may be more difficult than expected.

Besides the speed of sound, other quantities have been proposed to identify the presence of quark matter inside NS: the polytropic index $\gamma=d \mbox{ln}P/d\mbox{ln}\epsilon$, where $p$ is the pressure and $\epsilon$ is the energy density of the EoS \citep{Annala2019}, the trace anomaly $\Delta=1/3-P/\epsilon$ \citep{Fujimoto:2022ohj}, or a composition of these quantities, $d_c$, as defined in \cite{Annala:2023cwx}. Within a pQCD calculation these quantities take the values $c_s^2\lesssim 1/3$, $\gamma\, \in [1:1.7]$ and $\Delta\in [1:0.15]$ \citep{Annala:2023cwx}, at densities above 40$\rho_0$. In the conformal limit $\gamma=1$ and $\Delta=0$. 

  Since these quantities can characterize properties of pure nucleonic EoS, built from a RMF description, we calculate them with the 21 EoS discussed in the present work.  We have plotted $c_s^2$, $\gamma$ and $\Delta$ in Fig. \ref{fig:cs2x} for all the EoS.  We plot these quantities up to a density of the order of 1~fm$^{-3}$, which is approximately the central density of our maximum mass star, i.e. of stars with a mass $M_{TOV}$. At these densities, $\gamma$ takes values between 1 and 1.5, and $\Delta$ may be slightly positive, but in general it takes  negative  values, that can be as low as -0.2, in agreement with the NS data plotted in  Fig. 2 of \cite{Fujimoto:2022ohj} with results from \cite{Fujimoto:2017cdo,Fujimoto:2019hxv}. 

Finally, in Fig.~\ref{fig:dc}, we also plot the quantity proposed in \cite{Annala:2023cwx} as identifying the presence of quark matter,
$d_c=\sqrt{\Delta^2+\Delta'^2}$ with $\Delta'=c_s^2(1/\gamma -1)$. { $d_c=0.2$ would represent the threshold limit for the presence of quark matter: $d_c>0.2$ would imply purely hadronic matter while the $d_c<0.2$ would imply only quark matter.} While for some EoS, $d_c$ stays well above 0.2, there are some for which in the center of the star $d_c<0.2$.  The idea is that $\Delta$ and $\Delta'$ should be small when $\gamma$ and $c_s^2$ are close to the conformal limit. This would identify a phase transition to deconfined quark matter, which should have approximate conformal symmetry. We are able to get small values for $d_c$ because the polytropic index $\gamma$ takes quite small values, even though the speed of sound square never comes below 0.4. The models that predict a small $d_c$ are the ones with the smallest speed of sound square in the center of maximum mass star, $c_s^2\lesssim 0.5$, and these are the models with the largest contribution from the $\omega^4$ term, responsible for softening the EoS at high densities.

\subsection{Discussion}

In table \ref{tab:5} we summarize the behaviour of the 21 EoS with respect to several constraints that were discussed in previous subsections. These constraints include: the NS Mass-Radius from GW170817 within 50\% CI and 90\% CI, as well as the tidal deformability from the same event for mass ratio \( q = 1 \) or  NS Mass 1.36 $M_\odot$;  symmetry energy constraints from IAS;  pQCD  constraints  at 5 \( n_s \) and 7 \( n_s \) (\( n_s = 0.16 \) fm\(^{-3}\), and for renormalization scale \( X = 2 \)) from \cite{Komoltsev:2021jzg}; pure neutron matter constraint derived from $\chi$EFT calculations of the energy per particle from \cite{Huth22} between 0.04-0.08 fm\(^{-3}\) and 0.08-0.16 fm\(^{-3}\) density. EOS that satisfy the criteria are indicated with right tick marks, and those that do not meet the criteria are marked with cross ticks in the table. Only EoS 8, 18 and 19 satisfy all constraints. EoS 11, 13, 14, and 17 just do not satisfy the $\chi$EFT in the range 0.04-0.08 fm$^{-3}$, and EoS 16 fails the $\chi$EFT in all the ranges considered. EoS 6, 7, 9, 10, 12, 15 do not fall inside the E/N PNM envelop, and do not satisfy the IAS constraint. EoS 21 fails the tidal deformability and the pQCD at 7 $n_{s}$ constraints.

\begin{table*}
\caption{List of each EOS satisfying constraints on NS Mass-Radius from GW170817 within 50\% CI and 90\% CI, tidal deformability for mass ratio \( q = 1 \) or for NS Mass 1.36 $M_{\odot}$. Included are symmetry energy constraints from IAS, pQCD derived constraints at 5 \( n_s \) and 7 \( n_s \) (\( n_s = 0.16 \) fm\(^{-3}\), and the renormalization scale \( X = 2 \)), and pure neutron matter constraint derived from $\chi$EFT between 0.04-0.08 fm\(^{-3}\) and 0.08-0.16 fm\(^{-3}\). EOS that satisfy the criteria are indicated with right tick marks, and those that do not meet the criteria are marked with cross ticks in the table. \label{tab:5}}
\setlength{\tabcolsep}{7.5pt}
      \renewcommand{\arraystretch}{1.2}
\begin{center}    
\begin{tabular}{cccccccccc}
\hline
\multirow{2}{*}{EOS} & \multirow{2}{*}{Named} & \multicolumn{2}{c}{M - R GW} & $\Lambda_{1.36}$ & $S(\rho)$    & \multicolumn{2}{c}{pQCD X=2} & \multicolumn{2}{c}{E/N PNM envelop}                           \\ \cline{3-10} 
                     &                       & 50 CI         & 90 CI        & GW               & IAS          & 5 ns          & 7 ns         & $\rho = 0.04 -0.08$ fm$^{-3}$ & $\rho = 0.08 -0.16$ fm$^{-3}$ \\ \hline
EOS 1                &                       & $\tixzxmark$  & $\tixzxmark$ & $\tixzxmark$     & $\checkmark$ & $\tixzxmark$  & $\tixzxmark$ & $\checkmark$                  & $\checkmark$                  \\
EOS 2                &                       & $\tixzxmark$  & $\checkmark$ & $\tixzxmark$     & $\checkmark$ & $\checkmark$  & $\checkmark$ & $\tixzxmark$                  & $\tixzxmark$                  \\
EOS 3                &                       & $\tixzxmark$  & $\checkmark$ & $\tixzxmark$     & $\checkmark$ & $\checkmark$  & $\checkmark$ & $\tixzxmark$                  & $\checkmark$                  \\
EOS 4                &                       & $\tixzxmark$  & $\checkmark$ & $\tixzxmark$     & $\tixzxmark$ & $\checkmark$  & $\checkmark$ & $\tixzxmark$                  & $\tixzxmark$                  \\
EOS 5                &                       & $\tixzxmark$  & $\checkmark$ & $\tixzxmark$     & $\tixzxmark$ & $\checkmark$  & $\checkmark$ & $\tixzxmark$                  & $\checkmark$                  \\
EOS 6                &                       & $\checkmark$  & $\checkmark$ & $\checkmark$     & $\tixzxmark$ & $\checkmark$  & $\checkmark$ & $\tixzxmark$                  & $\tixzxmark$                  \\
EOS 7                &                       & $\checkmark$  & $\checkmark$ & $\checkmark$     & $\tixzxmark$ & $\checkmark$  & $\checkmark$ & $\tixzxmark$                  & $\tixzxmark$                  \\
{\bf EOS 8}                & UCIa                 & $\checkmark$  & $\checkmark$ & $\checkmark$     & $\checkmark$ & $\checkmark$  & $\checkmark$ & $\checkmark$                  & $\checkmark$                  \\
EOS 9                &                       & $\checkmark$  & $\checkmark$ & $\checkmark$     & $\tixzxmark$ & $\checkmark$  & $\checkmark$ & $\tixzxmark$                  & $\tixzxmark$                  \\
EOS 10               &                       & $\checkmark$  & $\checkmark$ & $\checkmark$     & $\tixzxmark$ & $\checkmark$  & $\checkmark$ & $\tixzxmark$                  & $\tixzxmark$                  \\
EOS 11               & UCIIa                & $\checkmark$  & $\checkmark$ & $\checkmark$     & $\checkmark$ & $\checkmark$  & $\checkmark$ & $\tixzxmark$                  & $\checkmark$                  \\
EOS 12               &                       & $\checkmark$  & $\checkmark$ & $\checkmark$     & $\tixzxmark$ & $\checkmark$  & $\checkmark$ & $\tixzxmark$                  & $\tixzxmark$                  \\
EOS 13               & UCIIb                & $\checkmark$  & $\checkmark$ & $\checkmark$     & $\checkmark$ & $\checkmark$  & $\checkmark$ & $\tixzxmark$                  & $\checkmark$                  \\
EOS 14               & UCIIc                & $\checkmark$  & $\checkmark$ & $\checkmark$     & $\checkmark$ & $\checkmark$  & $\checkmark$ & $\tixzxmark$                  & $\checkmark$                  \\
EOS 15               &                       & $\checkmark$  & $\checkmark$ & $\checkmark$     & $\tixzxmark$ & $\checkmark$  & $\checkmark$ & $\tixzxmark$                  & $\tixzxmark$                  \\
EOS 16               &                       & $\checkmark$  & $\checkmark$ & $\checkmark$     & $\checkmark$ & $\checkmark$  & $\checkmark$ & $\tixzxmark$                  & $\tixzxmark$                  \\
EOS 17               & UCIId                & $\checkmark$  & $\checkmark$ & $\checkmark$     & $\checkmark$ & $\checkmark$  & $\checkmark$ & $\tixzxmark$                  & $\checkmark$                  \\
{\bf EOS 18}               & UCIb                 & $\checkmark$  & $\checkmark$ & $\checkmark$     & $\checkmark$ & $\checkmark$  & $\checkmark$ & $\checkmark$                  & $\checkmark$                  \\
{\bf EOS 19}               & UCIc                 & $\checkmark$  & $\checkmark$ & $\checkmark$     & $\checkmark$ & $\checkmark$  & $\checkmark$ & $\checkmark$                  & $\checkmark$                  \\
EOS 20               &                       & $\checkmark$  & $\checkmark$ & $\checkmark$     & $\tixzxmark$ & $\checkmark$  & $\tixzxmark$ & $\tixzxmark$                  & $\tixzxmark$                  \\
EOS 21               &                       & $\checkmark$  & $\checkmark$ & $\tixzxmark$     & $\checkmark$ & $\checkmark$  & $\tixzxmark$ & $\checkmark$                  & $\checkmark$                  \\ \hline
\end{tabular}
\end{center}  
\end{table*}

\section{Conclusions}
We present twenty one EoS from the dataset generated in \cite{Malik:2023mnx}, that has been constrained to a few properties of nuclear matter (binding energy at saturation, saturation density, incompressibility), the neutron matter pressure calculated within a $\chi$EFT  approach, and a minimum 2$M_\odot$ NS maximum mass. The EoS were chosen in such a way that a large range of values of the slope of the symmetry energy at saturation is covered.

Unified inner-crust core EoS of $\beta$-equilibrium matter as expected inside neutron stars are built and several properties are discussed. The EoS satisfy the present available NS observations such as the ones from the GW170817 event \cite{LIGOScientific:2018cki}, NICER observations of the pulsar  PSR J0030 + 0451 \cite{Riley:2019yda,Miller:2019cac} and  radio and NICER observation of pulsar PSR J0740 + 6620 \cite{Fonseca:2021wxt,Riley:2021pdl,Miller:2021qha}. It is shown that although constrained by the $\chi$EFT pure neutron matter pressure calculations in the range 0.08-0.16 fm$^{-3}$  only half of them satisfy the pure energy per particle $\chi$EFT calculation of \cite{Huth22} in the same range of densities, and not all these EoS satisfy simultaneously the IAS constraints proposed in \cite{Danielewicz:2013upa}. 

Special attention has been set on the determination of the crust-core transition and the matching of the  crust to the core. We have analyzed the correlation of the slope of the symmetry energy at saturation with the crust-core transition density and proton fraction and proposed equations that translate these relations giving the corresponding absolute Pearson correlation coefficient.

Finally we have also discussed the behaviour of the speed of sound, trace anomaly and polytropic index, quantities frequently used to identify the possible existence of deconfined quark matter inside NS. It is shown that several of these EoS have in the center of the most massive NS a speed of sound squared of the order of $\lesssim 0.5$.  Most of the EoS predict a maximum central density of the order of 6$n_s$ with $n_s$ a reference saturation density of nuclear matter. Only three of the EoS do not satisfy the pQCD constraints discussed in \cite{Komoltsev:2021jzg} at 7$n_s$.

 Three of the EoS (8, 18 and 19) satisfy all the constraints discussed and four of them only fail the $\chi$EFT in the range 0.04-0.08 fm$^{-3}$ (EoS 11, 13, 14 and 17). The set that meets all the constraints is referred as UCIa to UCIc, and the one that fails only the lower part of the $\chi$EFT PNM constraints is referred as UCIIa to UCIId, see table~\ref{tab:5}. The designation UC stands for Universidade de Coimbra. Notice that although the M-R curves of these seven EoS are similar, predicting similar NS properties, the high density properties and the NMP vary, and in particular, the larger values of the incompressibility are associated with smaller symmetry energy properties.
EoS 16 fails the $\chi$EFT in all the ranges of densities considered. Six EOS (6, 7, 9, 10, 12, and 15), besides failing the $\chi$EFT constraint, also fail the IAS constraint in the complete range of densities given in \cite{Danielewicz:2013upa}. EoS 21 only fails two constraints: the tidal deformability for NS having mass 1.36 $M_\odot$ and pQCD at 7$n_{s}$.

\section{Data Availability}
All the calculated unified EOSs, together with their spectral fit for the inner core, will be available in  the platform \cite{compose}.

\section*{ACKNOWLEDGMENTS} 
This work was partially supported by national funds from FCT (Fundação para a Ciência e a Tecnologia, I.P, Portugal) under projects 
UIDB/04564/2020 and UIDP/04564/2020, with DOI identifiers 10.54499/UIDB/04564/2020 and 10.54499/UIDP/04564/2020, respectively, and the project 2022.06460.PTDC with the associated DOI identifier 10.54499/2022.06460.PTDC. HP acknowledges the grant 2022.03966.CEECIND (FCT, Portugal) with DOI identifier 10.54499/2022.03966.CEECIND/CP1714/CT0004. The authors acknowledge the Laboratory for Advanced Computing at the University of Coimbra for providing {HPC} resources that have contributed to the research results reported within this paper, URL: \hyperlink{https://www.uc.pt/lca}{https://www.uc.pt/lca}.

\bibliographystyle{aa}

\end{document}